%% file: main.tex
\documentstyle[epsfig,aps]{revtex}

\begin{document}
\draft

\title{\Large\bf Non-Linear Poisson-Boltzmann Theory of a Wigner-Seitz Model
for Swollen Clays}

\author{R. J. F. Leote de Carvalho$^{1}$,
E. Trizac$^{2}$ and 
J.-P. Hansen$^{3}$}

\address{
$ ^1$Department of Physics and Astronomy, University College London, \\
Gower Street, London WC1E 6BT, UK \\
and Laboratoire de Physique, Ecole Normale Sup\'erieure de Lyon,
(URA CNRS 1325)\\
46, All\'ee d'Italie, 69364 Lyon Cedex 07, France \\
$ $ \\
$ ^2$Laboratoire de Physique Th\'eorique, UMR 8627 du CNRS,\\ 
Universit\'e Paris XI, B\^atiment 210, 91405 Orsay Cedex, France \\
$ $ \\
$ ^3$Department of Chemistry, University of Cambridge \\
Lensfield Road, Cambridge CB2 1EW, UK}

\maketitle

\vspace{1cm}
\vspace{1cm}

\centerline{\bf Abstract}

Swollen stacks of finite-size disc-like Laponite clay platelets
are investigated within a Wigner-Seitz cell model. Each cell is a cylinder
containing a coaxial platelet at its centre, together with an overall 
charge-neutral distribution of microscopic co and counterions, within
a {\it primitive model} description. The non-linear Poisson-Boltzmann
(PB) equation for the electrostatic potential profile is solved 
numerically within a highly efficient Green's function formulation. 
Previous predictions of linearised Poisson-Boltzmann (LPB)
theory are confirmed at a qualitative level, but large quantitative 
differences between PB and LPB theories are found at physically relevant 
values of the charge carried by the platelets. A hybrid theory
treating edge effect at the linearised level yields good potential
profiles. The force between two coaxial platelets, calculated within
PB theory, is an order of magnitude smaller than predicted by LPB theory

\section{Introduction}
\label{sec:intro}

Electrostatic interactions between suspended mesoscopic particles or 
polyions play a key role in determining the stability, mesostructure and 
phase behaviour of colloidal dispersions and polyelectrolytes. Polyions in 
aqueous dispersions may be rigid or flexible, and range in shape from 
spherical to rod-like or lamellar. Widely studied examples of polyions
include polystyrene balls, elongated TMV (Tobacco Mosaic Virus) particles,
stiff polyelectrolyte chains like DNA, flexible membranes and silicate clay 
platelets. The highly charged polyions strongly attract or repel microscopic
counterions and coions (microions), leading to the formation of electric 
double layers characterised by highly inhomogeneous charge distributions 
around the polyions. Theoretical investigations of the structure of such 
double layers, and of their mutual effective interactions, have been mostly 
restricted to the simplest polyion topologies, including uniformly charged 
infinite planes or spheres. Starting with the pioneering work of Gouy
\cite{REF1} and Chapman \cite{REF2}, on infinite planar double-layers, 
theoretical calculations are almost invariably based on the mean-field
Poisson-Boltzmann (PB) approximation, which neglects excluded volume
and Coulomb correlations between the microions, although some recent attempts
have been made to include such correlations within a density functional
(DFT) formulation, which may also account for discrete solvent effects
\cite{REF3}. The non-linear PB equation for the local electrostatic 
potential may be solved analytically for a single planar double-layer, 
and two interacting double-layers in the salt-free case (i.e.
in the absence of coions) \cite{REF1,REF2,REF4}; in the presence of salt, 
the one-dimensional problem of interacting charged planes is easily solved 
numerically \cite{REF5}. However, PB theory becomes increasingly difficult 
to handle for more complicated geometries, and must generally be linearised 
to become tractable. A well-known application of linearised Poisson-Boltzmann 
(LPB) theory is the calculation of the effective interaction between
colloidal spheres carrying a uniform surface charge, leading to the 
screened Coulomb potential of Derjaguin, Landau, Verwey and Overbeek
\cite{REF6,REF7} (DLVO). LPB theory is, however, strictly speaking valid
only provided the local potential energy felt by the microions is 
everywhere small compared to the thermal energy $k_{_{B}} T$.
This condition is rarely met in the immediate vicinity of the 
highly-charged polyions, where the Coulomb energy becomes large 
compared to $k_{_{B}} T$, so that non-linearities become crucial and 
full PB theory should be used to determine the concentration 
profiles of the microions. The deficiencies of LPB theory are frequently
{\it patched up} by introducing the rather vague concept of {\it counterion
condensation}, leading to a Helmholtz-Stern layer of counterions highly 
bound to the polyions \cite{REF8}, and henceforth to an effective 
valence $Z_{eff}$ of the latter, significantly reduced in magnitude
compared to the nominal valence $Z$. It should be stressed that {\it 
counterion condensation} is a well-defined concept only within the 
cylindrical geometry of an infinite, uniformly charged thin rod
\cite{REF9,REF10}. In all other geometries, effective valences $Z_{eff}$ 
can only be defined within some phenomenological convention
\cite{REF11,REF12}. In the case of spherical charge-stabilised
colloids some recent direct measurements of the effective pair potential
between polyions in the bulk of a suspension point to the validity of 
the functional form and range of the DLVO potential, provided $Z_{eff}$ 
is considered to be an adjustable parameter, varying within a physically 
reasonable range \cite {REF13}.

This paper deals with electric double-layers in lamellar stacks of 
uniformly charged finite platelets, considered as a model for swollen
clays. Due to the finite size of the platelets, edge effects come
into play so that the problem ceases to be one-dimensional, as would be
the case for stacks of infinite planes. For simplicity, the clay
platelets are assumed to be disc-shaped and coaxial, while they are
equally spaced within an infinite stack. Fairly monodisperse 
disc-shaped platelets are synthetised as Laponite \cite{REF14},
a model clay which has received much recent experimental and theoretical
attention. In particular a Wigner-Seitz (WS) model for such stacks
has been thoroughly investigated within LPB theory \cite{REF15,REF16}.
Under most physical circumstances, the conditions of validity of LPB 
theory are far from being met in swollen clays, and consequently the 
much more difficult problem of non-linear PB theory for stacks of coaxial 
charged discs is being addressed in this paper.

The key innovation of the present work is that the solution of the non-linear
partial differential equation for given boundary 
conditions is not sought within
a standard finite difference scheme \cite{REF17}, but rather by expressing 
the solution in terms of the appropriate Green's function. The latter is 
obtained analytically for the present cylindrical geometry, in the form
of a Bessel-Dini series, and the resulting non-linear integral equation
is solved numerically by a very stable iterative procedure.

The remainder of the paper is organised as follows. The Wigner-Seitz model,
appropriate for stacks of coaxial charged discs, is defined in section II.
The basic assumptions and equations of PB theory are laid out in section
III, and the Green's function methodology for solving the non-linear PB 
equation with cylindrical WS boundary conditions is presented in section IV.
A hybrid PB/Debye-H\"uckel theory, whereby edge effects are treated to
linear order in the deviation of the potential from its counterpart for
infinite planar geometry, is formulated in section V. Numerical results for
potential and concentration profiles, and for the resulting osmotic 
properties, are presented in section VI. The force acting between two 
finite, coaxial platelets is calculated in section VII, while conclusions
are drawn in section VIII. A preliminary account of parts of this work was
published elsewhere \cite{REF18}.

\section{Wigner-Seitz model for swollen clays}
\label{sec:WS-model}

Consider a stack of coaxial, infinitely thin, disc-shaped clay
platelets, with an average spacing $H=2h$. Each platelet has
a radius $r_0$, and carries a uniform charge density 
$\sigma = -Ze/ \pi r_0^2$, where $Ze$ is the total charge on
a platelet. For Laponite, $r_0 \approx 150$\AA$\,$and $Z \approx 1000$
\cite{REF14}.
Stacks are assumed to fill space in a columnar arrangement, with the 
normals to the platelets all pointing in the same direction. Each 
platelet is then placed, on average, at the centre of a Wigner-Seitz cell,
of volume $v = V/N$, where $V$ is the total volume of the sample and
$N$ the total number of platelets. In a hexagonal columnar array, the
topology of the WS cell would be a prism of height $H$ and hexagonal
basis parallel to the platelet. 
The WS model thus describes a regular three-dimensional array of
platelets with perfectly aligned axes, and is thus appropriate to describe
moderately swollen stacked clays, but would not be appropriate for dilute
dispersions of clay particles.
In view of the circular shape of the 
platelet it is reasonable to replace the WS prism by a cylinder
of identical volume and height. In fact it was shown explicitly
in Ref. \cite{REF15} that, at least within LPB theory, the results
are very insensitive to the precise shape of the WS cell.
Compared to previous studies of stacked clay
platelets, assumed to be infinite planes \cite{REF5}, 
the present study takes into
account edge (or finite platelet size) effects.

The WS model for a swollen clay which will be examined in this paper
reduces hence to a single charged circular platelet placed at the 
centre of a coaxial cylinder, together with $N_+$ and $N_-$
monovalent counterions and coions. Overall charge neutrality
requires that
\begin{equation}
\label{electroneutrality}
N_+ = N_- + Z.
\end{equation}
The molecular nature of the solvent (water) inside the cell will be
ignored, i.e. the solvent is treated as a continuum of
dielectric constant $\epsilon$ ({\it primitive } model of electrolytes).
The height $H$ and radius $R>r_0$ of the cylinder must be such that 
$\pi R^2 H = v$, where the volume $v$ is determined by the macroscopic
concentration of clay platelets. The latter does not determine the aspect
ratio $H/R$, which will be determined by the condition that it minimizes
the total free energy of the microion distribution within the cell.
The task is now to determine the total electrostatic potential
$\varphi({\bf r})$ throughout the cell, and to derive from it the concentration 
profiles $\rho^+({\bf r})$ and $\rho^-({\bf r})$ of counterions and
coions, and the resulting osmotic properties, including the free energy,
for a given volume $v$ and aspect ratio $H/R$. In keeping with the physical
meaning of the WS cell as representing the cage formed by the 
neighboring platelets, the component of the electric field
{\bf E} normal to the surface $\Sigma$ of the cylinder will
be assumed to vanish at each point of that surface, i.e.
the following Neumann boundary condition will be imposed
throughout
\begin{equation}
\label{BC}
\left[ {\bf n}({\bf r}) \!\cdot\! 
\bbox{\nabla}\right]_{{\bf r}\in \Sigma} 
\varphi({\bf r})= 0,
\end{equation}
where ${\bf n}({\bf r})$ is the normal to the surface $\Sigma$ at ${\bf r}$.
The task thus defined will be carried out within PB theory in the following
sections.

\section{Poisson-Boltzmann theory.}
\label{sec:PB_theory}

The electrostatic potential throughout the WS cell satisfies Poisson's 
equation
\begin{equation}
\label{Poisson}
\nabla^2 \varphi({\bf r}) = - \frac{4 \pi} {\epsilon}  \rho_c({\bf r}),
\end{equation}
where $\rho_c({\bf r})$ is the total charge density, which includes the 
contribution of the platelet, $q_P({\bf r})$, and that of the 
(monovalent) microions
\begin{equation}
\label{ions_density}
\rho_c({\bf r}) = q_P({\bf r}) + e\left[\rho^+({\bf r}) - \rho^-({\bf r})\right].
\end{equation}
In the mean-field approximation, the positions of microions are
uncorrelated, so that their local concentrations are simply related
to the local potential by the following Boltzmann distributions
\begin{equation}
\label{Boltzmann}
\rho^\pm({\bf r})=\rho^\pm_0 \, {\rm e}^{\mp \beta e \varphi({\bf r})},
\end{equation}
where $\beta \equiv 1/k_{_{B}} T$. It is worth remembering that
Eq. (\ref{Boltzmann}) is just a statement of the Euler-Lagrange equation
resulting from the minimisation of the following free energy functional
\begin{eqnarray}
\label{functional_helmholtz}
{\cal F} \left[ \rho^+({\bf r}), \rho^-({\bf r}) \right] & = &
{\cal F}_{ideal} + {\cal F}_{Coulomb}
\nonumber \\
{\cal F}_{ideal} & = & \sum_{\alpha=+,-} k_{_{B}} T 
\int_v \rho^\alpha ({\bf r}) \left[ \ln \left( 
\Lambda_\alpha^3 \rho^\alpha({\bf r}) \right) - 1 \right]
\, d{\bf r}
\nonumber \\
{\cal F}_{Coulomb} & = & \frac{1}{2} \int_v \rho_c({\bf r})
\varphi({\bf r}) \, d{\bf r},
\end{eqnarray}
keeping in mind that $\rho_c({\bf r})$ and $\varphi({\bf r})$ are linearly
related by Poisson's equation.

The prefactors $\rho^\pm_0$ in Eq. (\ref{Boltzmann}) have no physical 
significance when considered separately. However, their product 
$\rho^+_0 \rho^-_0$ is uniquely determined by the imposed physical 
conditions. In a canonical description, $N_+$ and $N_-$ are fixed
by the macroscopic concentrations $n_+ \equiv N_+ / v$ and
$n_- \equiv N_- / v$ of counterions and coions; the latter coincides 
with the concentration $n_S$ of added salt, i.e. $n_- = n_S$.
The electroneutrality condition (\ref{electroneutrality}) then
fixes $n_+$, and the prefactors $\rho_0^\pm$ are determined by
the normalisation constraints
\begin{equation}
\label{concentrations}
n_\pm = \frac{1}{v} \int_{v} \rho^\pm({\bf r}) d{\bf r}.
\end{equation}
Alternatively, if the suspension is in osmotic equilibrium with an ionic 
solution of concentration $n_S'$ acting as a reservoir, a semi-grand-canonical
description is in order. In this case the chemical potential of co and 
counterions must be identical in the suspension and in the reservoir, leading to
the condition \cite{REF5}
\begin{equation}
\label{GC_condition}
\rho_0^+ \rho_0^- = \left(n'_S\right)^2.
\end{equation}
A key quantity describing the solution is then the Donnan ratio
$n_s/n'_S$ of salt concentrations in the cell and in the reservoir.
Substitution of Eq. (\ref{Boltzmann}) into Eqs. (\ref{ions_density}) and 
(\ref{Poisson}) leads to the closed non-linear partial differential
equation for the potential $\varphi({\bf r})$
\begin{equation}
\label{PB}
\nabla^2 \varphi({\bf r}) = -\frac{4 \pi}{\epsilon} 
\left( q_P({\bf r}) + e \left[ 
  \rho_0^+ {\rm e}^{-\beta e \varphi({\bf r})}
- \rho_0^- {\rm e}^{ \beta e \varphi({\bf r})} \right] \right).
\end{equation}
This PB equation must be solved subject to the Neumann boundary condition
(\ref{BC}) on the surface of the WS cylinder. The linearised version
of Eq. (\ref{PB}), where the Boltzmann factors are expanded to first
order in $\varphi$ (LPB theory) was solved analytically
in Refs. \cite{REF15,REF16}, under the same boundary conditions. 
The non-linear problem is re-formulated in terms of the
appropriate Green's function in the next section.

\section{Green's function methodology}
\label{sec:green}

The infinite dilution PB problem of an isolated platelet immersed with 
counterions in an electrolyte solution has been solved numerically
using a finite difference scheme \cite{REF17}.
However, this method is not well adapted to the case of a charged platelet 
confined to a WS cell. Here a semi-analytic method is proposed, whereby
the PB equation is transformed into a non-linear integral equation,
which can be solved iteratively. This approach requires the 
knowledge of a Green's function $G(\bf{r}, \bf{r}')$ such that 
its ``{\it convolution}'' product with the charge density $\rho_c({\bf r})$,
defined by Eqs. (\ref{Boltzmann}) and (\ref{ions_density}), namely
\begin{equation}
-\frac{4 \pi}{\epsilon}(G * \rho_c)({\bf r}) \equiv 
-\frac{4 \pi}{\epsilon}\int_v G({\bf r},{\bf r}') \, \rho_c({\bf r}') 
\,d{\bf r}'
\label{eq:convo}
\end{equation}
satisfies Poisson's equation (\ref{Poisson}) and the Neumann boundary condition
(\ref{BC}); $(G * \rho_c)$ coincides then with the required potential
$\varphi({\bf r})$. Note that the integral in Eq. (\ref{eq:convo})
is not, strictly speaking, a convolution since $G$ is not a function of the 
relative position ${\bf r}-{\bf r}'$ only.

By definition, the Green's function is a solution of the following 
{\bf linear} Poisson's equation
\begin{equation}
\nabla^2_{\bf r}\, G({\bf r}, {\bf r}') = \delta({\bf r} - {\bf r}').
\label{eq:green}
\end{equation}
Integrating both sides of Eq. (\ref{eq:green}), and using Gauss' theorem,
\begin{eqnarray}
\int_v \nabla_{\bf r}^2\, G({\bf r},{\bf r}') \, d{\bf r} 
& = & \int_{\Sigma}  {\bf n}({\bf r})  \!\cdot\! \bbox{\nabla}_{\bf r} 
G({\bf r},{\bf r}') \, dS
\nonumber \\
& = & \int_v \delta({\bf r} - {\bf r}') \, d{\bf r} = 1,
\end{eqnarray}
it becomes clear that the Green's function satisfying (\ref{eq:green})
cannot obey the same boundary condition (\ref{BC}) as the potential, 
but rather
\begin{equation}
\label{newBC}
\left[ {\bf n}({\bf r}) \!\cdot\! 
\bbox{\nabla}\right]_{{\bf r}\in \Sigma} 
G({\bf r},{\bf r}')= \frac{1}{{\cal S}(\Sigma)}
\end{equation}
where ${\cal S}(\Sigma)$ is the total area of the WS cell. The boundary
condition (\ref{newBC}) precludes the possibility of expanding $G$ in a 
Bessel-Dini series \cite{REF19}, similar to that used for the solution
of the LPB problem in the same cylindrical geometry \cite{REF15,REF16}.
The difficulty may be overcome by adopting one of three possible routes.
\begin{itemize}
\item[{\bf a)}] Eq. (\ref{eq:green}) is solved subject to the boundary condition
\begin{equation}
\left[ {\bf n}({\bf r}) \!\cdot\! 
\bbox{\nabla}\right]_{{\bf r}\in \Sigma} 
\left(G * \rho_c \right) ({\bf r}) = 0.
\end{equation}
An explicit example is given in Appendix A.
\item[{\bf b)}] A modified Green's function $G^{\cal B}$ is sought, which
satisfies the following Poisson's equation
\begin{equation}
\nabla^2_{\bf r}\, G^{\cal B}({\bf r}, {\bf r}') 
= \delta({\bf r} - {\bf r}') + {\cal B}(\bf r),
\label{eq:greenback}
\end{equation}
where ${\cal B}(\bf r)$ is an arbitrary neutralising background charge 
distribution, such that
\begin{equation}
\int_v {\cal B}\,d{\bf r} = -1.
\end{equation}
$G^{\cal B}({\bf r}, {\bf r}')$ can now be required to satisfy the same 
boundary condition (\ref{BC}) as the potential. It is easily verified that 
$-4 \pi  (G^{\cal B}*\rho_c) /\epsilon $ indeed satisfies Eq. (\ref{Poisson}),
together with the appropriate Neumann boundary condition (\ref{BC}), since
\begin{eqnarray}
\nabla^2 \left[ -\frac{4 \pi}{\epsilon}(G^{\cal B}*\rho_c) \right] &=& 
-\frac{4 \pi}{\epsilon}\int_v
\nabla^2_{\bf r}\, G^{\cal B}({\bf r}, {\bf r}')
\, \rho_c({\bf r}') \, d{\bf r}'
\nonumber\\
&=& -\frac{4 \pi}{\epsilon}\rho_c({\bf r}) 
-\frac{4 \pi}{\epsilon}{\cal B}({\bf r})\, \int_v\, 
\rho_c({\bf r}') \, d{\bf r}'
\nonumber\\
&=& -\frac{4 \pi}{\epsilon}\rho_c({\bf r})
\label{eq:expliback}
\end{eqnarray} 
where overall charge neutrality of the WS cell has been used; similarly
\begin{equation}
\left[ {\bf n}({\bf r}) \!\cdot\! 
\bbox{\nabla}\right]_{{\bf r}\in \Sigma} \,(G^{\cal B}*\rho_c)({\bf r}) 
= \int_v \,\left\{\left[{\bf n}({\bf r})\!\cdot\! 
\bbox{\nabla}\right]_{{\bf r}\in \Sigma} 
G^{\cal B}({\bf r},{\bf r}')\right\}\,
\rho_c({\bf r}')\, d{\bf r}' = 0.
\end{equation}
in view of the boundary condition imposed on $G^{\cal B}$.
\item[{\bf c)}] The bare Laplace operator in Eq. (\ref{eq:green})
is replaced by a {\it dressed} or {\it screened} operator, and the solution
$G^\kappa$ of
\begin{equation}
(\nabla^2_{\bf r} - \kappa^2) \, G({\bf r}, {\bf r}') 
= \delta({\bf r} - {\bf r}')
\label{eq:greenkappa}
\end{equation}
can then be made to satisfy the Neumann boundary condition 
(\ref{BC}) for any non-zero value of the inverse length
$\kappa$. The solution of this linear problem may be obtained
analytically along the same lines as those leading to the
potential within LPB theory
\cite{REF15,REF16}, as sketched in Appendix A. Subtracting
$\kappa^2 \varphi({\bf r})$ from both sides of the non-linear 
PB equation (\ref{PB}), it is then straightforward to check
that the solution to that equation, subject to the proper
boundary condition (\ref{BC}), satisfies the non-linear integral
equation
\begin{equation}
\varphi({\bf r}) = - \frac{4 \pi}{\epsilon}
\int_v G^\kappa ({\bf r},{\bf r}') \left( q_P({\bf r'}) + e\left[
  \rho_0^+ {\rm e}^{-\beta e \varphi({\bf r'})}
- \rho_0^- {\rm e}^{ \beta e \varphi({\bf r'})} \right]
+  \kappa^2 \frac{\epsilon}{4 \pi} \, \varphi({\bf r}') \right) d{\bf r}'.
\label{eq:varphi}
\end{equation}
Since $G^\kappa ({\bf r},{\bf r}')$ is known analytically,
Eq. (\ref{eq:varphi}) may be solved numerically by an iterative procedure, 
starting from an initial guess for $\varphi({\bf r})$.
\end{itemize}

Details for the analytic solutions of Eqs. (\ref{eq:green}), 
(\ref{eq:greenback}) and (\ref{eq:greenkappa}), subject to the appropriate 
boundary conditions, are given in Appendix A. In practice the iterative procedure
was implemented numerically using route c). Using the results from Appendix A, 
the required Green's function may be written in the form of the following 
Bessel-Dini series
\begin{equation}
\label{green}
G^\kappa ({\bf r},{\bf r}') = 
\sum_{n \geq 1} {\cal{C}}^\pm_n (\phi,{\bf r}')
J_0\left(y_n \frac{r}{R}\right)
\cosh \left[ \frac{h\mp z}{\Lambda_n} \right],
\end{equation}
where ${\bf r} = (r, \phi,z)$ (cylindrical coordinates), 
the signs $+$ and $-$ correspond to the 
situations $z > z'$ and $z < z'$ respectively, 
$y_n$ is the n$^{th}$ root of $J_1(y)=0$,
$J_0$ and $J_1$ are the zeroth and first order cylindrical Bessel 
functions, and $\Lambda_n^{-2} = (y_n^2 + \kappa^2 R^2) / R^2$. 
The coefficients ${\cal{C}}^\pm_n(\phi,{\bf r}')$ are given by
\begin{equation}
{\cal{C}}^\pm_n(\phi,{\bf r}') = 
- 2 \Lambda_n \frac{\delta(\phi-\phi')}
{R^2 \sinh \left[2h/\Lambda_n \right]} 
\frac{J_0 \left( y_n\, r'/R \right)}{J_0^2(y_n)}
\cosh\left[\frac{h\pm z'}{\Lambda_n} \right].
\label{eq:cn}
\end{equation}
For an infinitely thin platelet of radius $r_0$, the charge density 
$q_P({\bf r})$, which provide the source term in Poisson's equation,
may be simply expressed as
\begin{equation}
\label{disc_charge_density}
q_P({\bf r}) = \sigma \, \Theta(r_0 - r) \,  \delta(z),
\end{equation}
where $\Theta$ is the Heaviside function. The corresponding
contribution to $\varphi({\bf r})$ in Eq. (\ref{eq:varphi}) 
may then be evaluated 
analytically
\begin{equation}
\label{analytical_source}
\int_v d{\bf r}' G^\kappa ({\bf r},{\bf r}') \, q_P({\bf r}') 
= - \sigma \frac{r_0}{R} \sum_{n \ge 1}^{\infty} \Lambda_n
\, \frac{ J_1 \left( y_n\, r_0/R \right) }
{ y_n \sinh \left[ h / \Lambda_n \right] }
\, \frac{ J_0 \left( y_n\, r/R \right) } { J_0^2(y_n) } \,
\cosh\left[ \frac{h \mp z}{\Lambda_n} \right].
\end{equation}
The non-linear integral equation (\ref{eq:varphi}) was then solved for
$\varphi({\bf r})$ using an iterative Picard method with 
underrelaxation \cite{REF20}. At a particular iteration step 
$i$, the result from the previous iteration, 
$\varphi^{i-1}({\bf r})$, and the analytic result 
(\ref{analytical_source}) are used as input to compute the r.h.s 
of Eq. (\ref{eq:varphi}). The resulting potential $\varphi({\bf r})$
is then used to produce the input for the next iteration step
by mixing it with $\varphi^{i-1}({\bf r})$ according to
\begin{equation}
\varphi^i({\bf r})=\alpha \, \varphi^{i-1}({\bf r}) 
+ (1-\alpha) \, \varphi({\bf r})
\end{equation}
where $0<\alpha<1$; typically $\alpha \approx 0.9$. The iterative
cycle is repeated until the relative difference 
$|\varphi^{i}-\varphi^{i-1}|/|\varphi^{i}|$ at the centre of the 
cylindrical WS cell (which coincides with the centre of the disc) 
becomes smaller than a pre-set value, chosen to be  $10^{-5}$
in practice.

The arbitrary inverse length $\kappa$ is chosen in the range
$[\kappa_{_{D}}/10,10\,\kappa_{_{D}}]$, where 
$\kappa_{_{D}} \equiv \left( 8 \pi n_S' e^2 / (\epsilon k_{_{B})} 
T \right)^{1/2}$ is the inverse Debye length in the reservoir, in 
the case of calculations carried out in the semi-grand-canonical 
ensemble. The prefactors $\rho_0^+$ and $\rho_0^-$ in Eq. 
(\ref{Boltzmann}) are fixed during the calculation at values
obeying the constraint (\ref{GC_condition}). Note that the 
electroneutrality condition (\ref{electroneutrality})
is not obeyed until the iterations have converged to the
actual solution $\varphi({\bf r})$, thus providing an additional
global convergence test. If the calculations are carried out in the 
canonical ensemble, $n_S = n_-$ is fixed, and  
$\kappa_{_{D}} \equiv [8 \pi n_S e^2 / (\epsilon k_{_{B}} T)]^{1/2}$.
The concentration of counterions is fixed by the electroneutrality
condition (\ref{electroneutrality})
and the prefactors $\rho_0^+$ and $\rho_0^-$ are determined by
Eq. (\ref{concentrations}) at each step of the iteration, so that
electroneutrality holds throughout the convergence process.

The converged solution $\varphi({\bf r})$ must be independent
of the particular choice of the auxiliary variable $\kappa$.
Cylindrical symmetry implies $\varphi({\bf r})=\varphi(r,z)$, which is 
calculated on a 2-dimensional grid $n_r \times n_z$ spanning half the 
cylinder. In practice $n_r=240$ grid points were used to cover
the interval $[0,R]$, which provides sufficient resolution for
the representation of the Bessel functions, and $n_z=100$ points for 
the interval $[0,h]$. 60 terms were retained in the Bessel-Dini series 
(\ref{green}). Starting from the initial guess $\varphi({\bf r}) \equiv 0$,
convergence is generally achieved in about 50 to 200 iterations.
The number of iterations needed to achieve a relative accuracy of
$10^{-5}$ can be drastically reduced if the electrostatic potential
obtained from the hybrid theory, introduced in the next section,
is used as input.

\section{A hybrid Poisson-Boltzmann/Debye-H\"uckel theory}
\label{sec:pbdh}

The hybrid Poisson-Boltzmann/Debye-H\"uckel (PB/DH) theory was developed 
for the problem of a clay platelet in a cylindrical WS cell
as a first attempt to go beyond LPB \cite{REF15}. 
Within this symmetry the potential $\varphi({\bf r})=\varphi(r,z)$ can 
be expanded in a Bessel-Dini series \cite{REF19}
\begin{equation}
\label{Bessel_Dini}
\beta e \, \varphi(z,r) = \sum^\infty_{n=1} A_n(z) 
J_0\left(\frac{y_n r}{R}\right),
\end{equation}
which factorizes the dependence of the potential on
the distance $z$ along the axis of the cylinder, and
the radial distance $r \equiv (x^2 + y^2)^{1/2}$. 
The hybrid PB/DH approach is an attempt to treat edge effects in a
perturbative way, whilst keeping the non-linear PB description in the
limit of infinite platelets.

Substituting (\ref{Bessel_Dini}) into Eq. (\ref{Boltzmann}), the leading 
term $A_1(z)$ 
in the expansion is exponentiated, but the exponential is linearised 
with respect to the remainder of the series ($n \geq 2$)
\begin{equation}
\label{semi_linearised_expansion}
\rho^\pm = \rho_0^\pm \, {\rm e}^{\mp A_1(z)}
\left( 1 \mp \sum_{n=2}^\infty
A_n(z) J_0 \left( y_n \frac{r}{R} \right) \right).
\label{eq:semiexpansion}
\end{equation}
The first term in this expansion, $n=1$, 
corresponds to the solution of the PB equation for
an infinite charged plane in a WS slab of height
$2h$, with an effective surface charge $\sigma' =
\sigma (r_0/R)^2$. The terms of order $n \ge 2$ may 
be considered as providing an estimate of the correction to
the infinite charged plane limit ($r_0 \to R$) 
due to the finite size of the platelets \cite{REF15}. 

The concentrations of co and counterions are determined by
\begin{equation}
\label{concentrations_PB.BH}
n^\pm = \rho_0^\pm \frac{1}{2h} \int^{+h}_{-h} 
e^{\mp A_1(z)} dz.
\end{equation}
It is therefore impossible to impose
$\rho_0^+=n^+$ and $\rho_0^-=n^-$ simultaneously.
As a consequence, in the presence of added salt, calculations 
are preferably performed in the semi-grand canonical ensemble, 
at fixed reservoir salt concentration. This can be achieved 
with the choice
\begin{equation}
\label{constraint_to_rho_0}
\rho_0^+=\rho_0^-=n_S'.
\end{equation}

For $z\ne 0$ the semi-linearised Poisson-Boltzmann equation
becomes
\begin{eqnarray}
\label{PB_DH_first}
\nabla^2 \varphi(\bf r) & = & -\frac{4 \pi e}{\epsilon} \times
\nonumber \\ & & \left\{ 
  \rho_0^+ \, {\rm e}^{- A_1(z)} 
- \rho_0^- \, {\rm e}^{ A_1(z)}
- \left(\rho_0^+ \, {\rm e}^{-A_1(z)} 
      + \rho_0^- \, {\rm e}^{ A_1(z)} \right)
\sum_{n \ge 2}^\infty A_n(z) J_0\left( y_n \frac{r}{R}\right) \right\}.
\end{eqnarray}
The boundary condition associated with the presence of the uniformly 
charged platelet at $z=0$, $r < r_0$ is
\begin{equation}
\left( \frac{\partial \varphi(z,r)}{\partial z}\right)_{z=0^+} = 
- \frac{2 \pi \sigma}{\epsilon} \Theta(r_0 - r).
\end{equation}
The electric field normal to the top and bottom of the cylinder vanishes, 
i.e.
\begin{equation}
\left( \frac{\partial \varphi(z,r)}{\partial z}\right)_{z=\pm h} = 0 .
\end{equation}
Using (\ref{constraint_to_rho_0}) in (\ref{PB_DH_first})
and then projecting on the basis of the zeroth order 
Bessel functions $J_0$ leads to the following set of 
differential equations
\begin{eqnarray}
\label{PB_DH_num}
\frac{d^2 A_1(z)}{dz^2} & = & 8 \pi l_B n_S' \sinh A_1(z) \nonumber \\
\frac{d^2 A_n(z)}{dz^2} - \left( \frac{y_n}{R} \right)^2 
A_n(z) & = & 8 \pi l_B n_S' A_n(z) \sinh A_1(z)
\makebox[0.5cm][r]{, }n\ge 2 
\end{eqnarray}
with the boundary conditions at $z=0$
\begin{eqnarray}
\left( \frac{d A_1(z)}{dz}\right)_{z=0^+} & = & - \frac{1}{b}
\left( \frac{r_0}{R} \right)^2 \nonumber \\
\left( \frac{d A_n(z)}{dz}\right)_{z=0^+} & = & - \frac{2}{b}
\left(\frac{r_0}{R}\right) \frac{1}{y_n J^2_n(y_n)}
J_1\left(\frac{y_n r_0}{R}\right)
\makebox[0.5cm][r]{, }n\ge2
\end{eqnarray}
and at $z= \pm h$
\begin{equation}
\left( \frac{d A_n(z)}{dz}\right)_{z=\pm h} = 0 
\makebox[0.5cm][r]{, }n\ge 1 
\end{equation}
where $l_B = \beta e^2 / \epsilon$ is the Bjerrum length and 
$b=e / 2 \pi \sigma l_B$ the Gouy length.

This set of differential equations can be solved analytically
in the salt-free case, for a WS cylindrical cell of 
infinite height \cite{REF15} ($h \to \infty$, $R$ finite).
A numerical solution must be sought otherwise.
The first equation in (\ref{PB_DH_num}) is identical to
the non-linear PB equation for the
ion distribution in a 1-dimensional cell model where each 
platelet is assumed to be an infinite uniformly charged plane 
confined with its monovalent co and counterions 
to a slab. In order to solve this equation
we follow the prescription given in Appendix B 
of \cite{REF5}. A first order differential
equation can be obtained after integrating once,
using the boundary condition at $z=h$
\begin{equation}
\frac{d A_1(z)}{dz} = - 
\sqrt{ 16 \pi l_B n_S' \left( \cosh A_1(z)- 
\cosh A_1(h)\right) }.
\end{equation}
The equation above is solved numerically using 
a 4th order Runge-Kutta algorithm. Different guesses for
$A_1(h)$ lead to different solutions
$A_1(z)$. An underestimated initial guess for
$A_1(h)$ is increased until
a solution $A_1(z)$ that verifies
the boundary condition at $z=0$ is found.

The second equation in (\ref{PB_DH_num}) can also 
be solved by first reducing it to a first order
differential equation and then using an appropriate
numerical algorithm for its integration.
Defining a function
\begin{equation}
L(z) \equiv 8 \pi l_B n_S' \sinh A_1(z)-
\left(\frac{y_n}{R}\right)^2,
\end{equation}
one obtains
\begin{equation}
\frac{d^2 A_n(z)}{dz^2} - L(z) A_n(z) = 0,
\end{equation}
which is a {\sl stiff} differential equation if
$L(z)$ is large. This can be reduced to a first
order differential equation by applying a Ricatti 
transform
\begin{equation}
\label{ricatti}
\eta(z) = \frac{1}{A_n(z)}\frac{d A_n(z)}{dz},
\end{equation}
which leads to
\begin{equation}
\frac{d\eta(z)}{dz}=L(z)-\eta(z)^2.
\end{equation}
The differential equation above is again solved 
using a 4th order Runge-Kutta
algorithm, and forced to satisfy the boundary 
condition at $z=h$ re-written as 
\begin{equation}
\eta(h)=0.
\end{equation}
Integration of (\ref{ricatti}) leads to
\begin{equation}
A_n(z) = C \exp\left[ \int_0^z \eta(z') dz'\right].
\end{equation}
The integral above can be evaluated numerically while
the integration constant $C$ is determined by
the boundary condition at $z=0$
\begin{equation}
C = \frac{1}{\eta(0)}\left(\frac{d A_n(z)}{dz}\right)_{z=0}.
\end{equation}

The Bessel-Dini coefficients $A_n(z)$ in the expansion
(\ref{Bessel_Dini}) of the electrostatic potential can be computed 
numerically using the procedure just described. This involves
solving as many independent differential equations as terms kept 
in the expansion, {\it ie} one for each
coefficient. The computational cost is considerably 
reduced compared to the solution of the non-linear 
PB equation using the iterative scheme prescribed in 
the previous section.

\section{Results}
\subsection{Potential and density profiles}
\label{sec:pot}

Figs. \ref{Potentials}a-c compare the dimensionless electrostatic
potential $\Phi(z,r) \equiv \beta e \varphi(z,r)$ as obtained within LPB 
\cite{REF15,REF16}, PB/DH and PB. The PB/DH potential 
was computed by first solving (\ref{PB_DH_num}) for the 
coefficients $A_n(z)$ and then using these in the Bessel-Dini expansion 
(\ref{Bessel_Dini}) for the electrostatic potential.
Only the projections $(z=0,r)$ and $(z,r=0)$ are shown. The results are 
for an aqueous solution of clay discs of radius $r_0 = 150$\AA$\,$at 
temperature $T = 300$K. The aspect ratio of the cylindrical WS cell is 
$h/r_0=1.25$ and the concentration of clay $n=5 \times 10^{-3}$M
(1M = 1 mol dm$^{-3}$). Although in the present 
model the clay platelets are assumed to be
infinitely thin, a packing fraction can be defined
by assining a finite thickness $d$ to the platelets; 
$d$ can be chosen to be 1nm,
as appropriate for Laponite \cite{REF14}.

The PB and PB/DH results in Fig. \ref{Potentials} are semi-grand canonical 
calculations for $n_S' = 5 \times 10^{-3}$M and $Z=5, 100$ 
and 1000. The Donnan ratios $n_S/n_S'$ in Table \ref{table1} compare
well, indicating that the concentration of added salt in the WS cell is
approximately the same in PB and PB/DH, for a fixed $Z$. Unlike other
thermodynamic quantities (see next section), this macroscopic property 
is well estimated within the PB/DH theory.

The LPB results of Fig. \ref{Potentials}
have been produced for the same salt concentrations $n_S$ in the cell
as those obtained within the semi-grand canonical PB theory 
(see Table \ref{table1}). Once $n_S$ is known,
the Donnan ratio reads within LPB \cite{REF21}
\begin{equation}
n_S/n_S' = (1+Zn/n_S)^{-1/2}.
\end{equation}
The corresponding data are collected in Table \ref{table1}.
The Donnan ratio in LPB is smaller that in PB and the difference
increases with $Z$. For a given salt concentration in the reservoir,
LPB theory underestimates the concentration of added salt 
in the WS cell.

Turn next to the comparison of the electrostatic potentials in
Fig. \ref{Potentials}. Although at very low surface charge 
($Z = 5$) all theories produce an identical electrostatic potential, 
as soon as $Z$ is as large as 100, it becomes obvious that the 
linearisation of the Boltzmann factors is no longer an acceptable 
approximation. However, at such high surface charge densities, the PB/DH still 
provides a very good approximation for $\varphi(z,r)$. At physically relevant 
charge densities, e.g. $Z=1000$, the LPB approximation gives rise to an 
electrostatic potential substantially different from that obtained within PB 
or PB/DH. The PB/DH theory still yields a good approximation for the
potential except near the surface of the platelet, for
$r \stackrel{>}{\sim}r_0$. This illustrates the breakdown of 
linearisation to account for finite size corrections. 

Up to a rescaling factor, the shapes of PB and LPB 
potentials are similar (Fig. \ref{Potentials}). For a given charge density
on the platelets (fixed $Z$), one can then define an effective charge 
$Z^*$ from a maximum likelyhood criterion between $\varphi^{PB}(Z)$ and 
$\varphi^{LPB}(Z^*)$: for $Z=1000$ and the parameters of Fig. \ref{Potentials}, 
the corresponding effective charge is $Z^*=375$ (see Fig. \ref{fig:renorm}). 
As expected, $Z^*<Z$, which means that the counterions recondense within a 
thin layer around the particle until the electric potential near the surface is 
lowered to a value of a few $k_{_{B}} T$. Fig. \ref{fig:renorm} shows that
the resulting agreement is only qualitative and that the renormalized LPB 
potential can neither account for the edge effects nor for the behaviour 
along the $z$ axis. However, in the vicinity of the WS surface and far 
from the platelet where the variations of the potential are weak, PB theory 
can be linearised and the potential retains an LPB form, provided the 
bare charge is replaced by an effective charge $Z^*$. In the simpler
spherical geometry, given the non-linear PB
electric potential, Alexander {\it et al.} \cite{REF11} proposed
to determine the effective parameters for the linearised theory
by matching to the far-field limit of the PB solution: the resulting
LPB potential, matching the PB potential up to its third derivative, 
is then an effective potential for the colloid-colloid interactions.
The phase diagram associated with the latter effective interactions
was later shown to be in good agreement with experiments for 
charged spherical colloids \cite{REF22}. In the case of non-spherical
colloids in anisotropic cells, the condensation of micro-ions
may well be non uniform so that the effective particle may have 
a different charge distribution than the original one. Preliminary
results confirm this scenario for clay platelets:
the effective charge $Z^*$ depends on the point chosen on the WS surface 
to match the LPB and PB  potentials, except of course when $Z$ is low enough.
The concept of charge renormalization thus seems to be inadequate
for clay particles due to the non uniformity of micro-ions condensation.

\subsection{Osmotic properties and quadrupole moment}
\label{sec:thermo}

Once the potential $\varphi({\bf r})$ and the concentration profiles 
$\rho^+({\bf r})$ and $\rho^-({\bf r})$ have been determined within
PB, PB/DH or LPB theories, a number of osmotic properties may be
calculated, as discussed in detail in Refs. \cite{REF15,REF16}. The
Helmholtz free energy is obtained by substituting the profiles
into the functional (\ref{functional_helmholtz}), with the result
\begin{equation}
\label{eq:helmholtz}
\beta F = \beta (U_P - U_C) + ( N^+ + N^- )
\left[ \ln \left(n'_S \lambda^3\right) -1 \right].
\end{equation}
where $\lambda$ is an irrelevant length scale, and
$U_P$ and $U_C$ are the following two contributions to the
internal energy $U=U_P + U_C$
\begin{equation}
U_P = \frac{1}{2} \int_v q_P({\bf r}) \varphi({\bf r}) \, d{\bf r},
\makebox[1.0cm][r]{ }
U_C = \frac{e}{2} \int_v \left[\rho^+({\bf r})-\rho^-({\bf r})
\right] \varphi({\bf r}) \, d{\bf r}.
\label{eq:internalen}
\end{equation}
Expression (\ref{eq:helmholtz}) is valid within the semi-grand
canonical ensemble (which was used throughout), with the choice
of $\rho_0^+ = \rho_0^- = n_S'$.
The resulting grand potential is
\begin{equation}
\beta \Omega = \beta F - 2 N^- \ln  \left( n'_S \lambda^3 \right).
\end{equation}
For a given clay concentration, and hence for a given WS volume
$v$, $\Omega$ and all osmotic properties depend on the aspect
ratio $h/R$ (or equivalently $h/r_0$). The equilibrium
topology of the columnar stacking of platelets,
as represented by the WS cell model, is that which minimizes
$\Omega$ as a function of $h/r_0$, for given values of $v$,
$n_S'$, $T$ and $\epsilon$. For convenience, the results shown
in the figures are shifted by  
$\Omega_0 \equiv Z \ln(n'_S \lambda^3)$, {\it ie}
\begin{equation}
\beta (\Omega - \Omega_0) = \beta (U_P - U_C) + N^- + N^+.
\end{equation}
The osmotic pressure $\Pi$ can be derived from the pressure tensor
\cite{REF15,REF16}
\begin{equation}
\Pi = \frac{1}{\beta} \sum_{k=+,-} \overline{ \rho^k }^\Sigma
+ \frac{\epsilon}{8 \pi}\overline{ (\bbox{\nabla} \varphi)^2 }^\Sigma,
\end{equation}
where $\overline{(..)}^{\Sigma}$ denotes an average over 
the total surface $\Sigma$
of the WS cell. The osmotic pressure 
in the reservoir is $\Pi_0 \equiv 2 \,k_{_{B}} T \,n'_S $.
For a cylindrical WS cell a disjoining pressure can be defined by
\begin{equation}
\label{eq:disjoining_pressure}
\Pi_d = \frac{1}{\beta} \sum_{k=+,-} \overline{ \rho^k }^{S'}
+ \frac{\epsilon}{8 \pi}\overline{ |\bbox{\nabla} \varphi|^2 }^{S'},
\end{equation}
where $S' \subset \Sigma$ is the surface of the bottom and top of 
the WS cell. The swelling arises from the osmotic pressure exerted by the
micro-ions and a normal spacing between the platelets 
(equal to the height of the cylindrical cell) is obtained when an axial
pressure 
$\Pi_d$ is applied.

The {\it a priori} lowest order non-vanishing multipole moment
of the charge distribution in the WS cell is the $zz$ component 
of the traceless quadrupole tensor
\begin{equation}
\label{Q_volume}
Q_{zz}^{tot} = \frac{1}{2} \int_v d{\bf r} \left[
\rho_c({\bf r}) \right] (2 z^2 - x^2 - y^2).
\label{eq:quadrudef}
\end{equation}
Upon integration by parts, this may be re-expressed as a 
surface integral
\begin{equation}
\label{Q_surface}
Q_{zz}^{tot} =  \frac{\epsilon}{4 \pi} \frac{1}{2} 
\oint_\Sigma \varphi({\bf r}) \bbox{\nabla}(2z^2 - x^2 - y^2) \cdot {\bf d S},
\label{eq:quadrusurface}
\end{equation}
where Poisson's equation has been used.
Numerical accuracy of the solution $\varphi({\bf r})$ to the PB equation
can be checked by comparing the quadrupole as obtained from the two 
equations (\ref{eq:quadrudef}) and ({\ref{eq:quadrusurface}). In the plots,
$Q_{zz}^{tot}$ is normalised by the quadrupole moment of the disc 
\begin{eqnarray}
Q_{zz}^{disc} & =& \frac{1}{2} \int_v q_P({\bf r})\, (2 z^2 - x^2 - y^2)\,
d{\bf r}
\nonumber \\
& = & \frac{Z e r_0^2}{4}.
\end{eqnarray}

Figs. \ref{PB.DH_comparisons}a-c show results for the osmotic
and disjoining pressures, grand potential and quadrupole moment
as functions of the aspect ratio $h/r_0$, for $Z=100$. Results
from PB and PB/DH are compared. Within the latter theory, the
concentration profiles may be calculated from Eq. 
(\ref{semi_linearised_expansion}), once the coefficients
$A_n$ have been obtained; alternatively, the profiles may be 
calculated by re-exponentiating the potential determined
from Eq. (\ref{Bessel_Dini}), according to Eq. (\ref{Boltzmann}).
Fig. \ref{PB.DH_comparisons}a shows that the osmotic pressure 
$\Pi$ goes through a minimum at a well-defined aspect ratio. 
Within PB theory, the disjoining pressure $\Pi_d$ coincides 
with the osmotic pressure at that minimum. The re-exponentiated 
version of PB/DH theory leads to results in reasonable agreement with the 
PB data, whereas straight PB/DH theory yields an unphysical osmotic
pressure which is lower than the reservoir pressure $\Pi_0$;
note that although the corresponding disjoining
pressure agrees well with the PB data, it intersects the osmotic
pressure curve away from the minimum in the latter, and at
a much higher aspect ratio. Fig. \ref{PB.DH_comparisons}b
shows that the PB results for the grand potential $\Omega$
go through a flat minimum at the same aspect ratio as the osmotic
pressure. The minimum is shifted to considerably higher aspect
ratios within PB/DH theory and its re-exponentiated form,
and does not coincide with the corresponding minima in
the osmotic pressure curves. The PB quadrupole moments calculated
from Eqs. (\ref{Q_volume}) and (\ref{Q_surface}) coincide and
go through zero at the value of $h/r_0$ corresponding to the
osmotic pressure and grand potential minima of the Figs
\ref{PB.DH_comparisons}a,b. The PB/DH results are fairly close 
to their PB counterparts but go through zero at an aspect ratio
which is significantly lower than the location of
the minimum of the corresponding grand potential.
The re-exponentiated PB/DH results, on the other hand, as 
calculated from Eqs. (\ref{Q_volume}) and (\ref{Q_surface})
are inconsistent.

Calculations at the very low platelet charge $Z=5$ lead to
undistinguishable results between the PB and re-exponentiated
PB/DH theories, as one might expect from the nearly identical
potential profiles in Fig. \ref{Potentials}. However, at the 
physically relevant charge $Z=1000$, the discrepancies between 
the two theories become very large, as illustrated in Fig.
\ref{PB.DH_vs_PB}.

The scenario emerging from PB theory is reminiscent of the
predictions of LPB theory. In the framework of the latter
the Helmholtz free energy goes through a minimum at the 
same value of the aspect ratio $h/r_0$ for which the 
disjoining and osmotic pressures are equal. 
The quadrupole moment $Q$ vanishes at this same 
$h/r_0$. The same behaviour follows from the present 
PB results in the semi-grand canonical ensemble, showing 
that $Q_{zz}^{tot}=0$ and $\Pi = \Pi_d$ at the aspect ratio 
that minimises the grand potential (see Figs. 
\ref{PB.DH_comparisons} and \ref{PB.DH_vs_PB}).
This illustrates the robustness of the equivalence between 
\begin{itemize}
\item a thermodynamic minimization
\item an electrostatic criterion ($Q_{zz}^{tot}=0$)
\item a mechanical equilibrium condition ($\Pi=\Pi_d$)
\item the minimization of an osmotic constraint: 
$\Pi$ is minimum at the aspect ratio where $\Pi=\Pi_d$ (see Fig.
\ref{PB.DH_comparisons}a and \ref{PB.DH_vs_PB}a. Also holds within LPB).
\end{itemize}
The equilibrium separation between two platelets is determined 
by the minimisation of the grand potential (or Helmholtz free energy). 

Being an approximate theory for the electrostatic potential, the 
PB/DH is inconsistent at $Z \stackrel {>}{\sim} 100$.
However, in spite of all the deficiencies which the PB/DH theory 
exhibits when trying to compute macroscopic quantities from the electrostatic 
potential, this approximation does manage to give a good first approximation 
for $\varphi({\bf r})$ which compares well with that obtained by solving 
the PB equation.

\section{Force acting between two parallel platelets}
\label{sec:force}

The Green's function methodology of section \ref{sec:PB_theory}
may be extended to calculate the potential and density profiles
around two coaxial platelets placed symmetrically inside a WS cell,
within PB theory. If the two uniformly charged parallel discs are
placed at $z= \pm z_0$ from the centre of a WS cell as shown in Fig.
\ref{fig2plates}, the PB equation is still given by 
(\ref{PB}), with the source 
term $q_P({\bf r})$ now including a contribution from both platelets
\begin{equation}
\label{new_source_term}
q_P({\bf r}) = \sigma \, \Theta(r_0 - r) \,
\left[ \delta(z+z_0)   + \delta(z-z_0)  \right].
\end{equation}
The boundary conditions (\ref{BC}) still applies and the integral
equation (\ref{eq:varphi}) for the local potential 
$\varphi({\bf r})$ remains unchanged, with the contribution 
(\ref{analytical_source}) from the source term now following from
(\ref{green}) and (\ref{new_source_term}), with the result
\begin{equation}
\int_v d{\bf r}' G^\kappa ({\bf r},{\bf r}') \, q_P({\bf r}') 
= - 2 \sigma \frac{r_0}{R} \sum_{n \ge 1}^{\infty} \Lambda_n
\, \frac{ J_1 \left( y_n\, r_0/R \right) }
{ y_n \sinh \left[ h / \Lambda_n \right] }
\, \frac{ J_0 \left( y_n\, r/R \right) } { J_0^2(y_n) }
\, \Upsilon(z),
\end{equation}
where the function $\Upsilon(z)$ is given by
\begin{equation}
\Upsilon(z) = \left\{ \begin{array}{ll}
\cosh\left[ z_0 / \Lambda_n \right] \,
\cosh\left[ (h-z) / \Lambda_n \right]
\makebox[1.0cm][r]{ },\,z_0<z \\
\cosh\left[ z / \Lambda_n \right] \,
\cosh\left[ (h-z_0) / \Lambda_n \right]
\makebox[1.0cm][r]{ },\,-z_0<z<z_0 \\
\cosh\left[ z_0 / \Lambda_n \right] \,
\cosh\left[ (h+z) / \Lambda_n \right]
\makebox[1.0cm][r]{ },\,z<-z_0.
\end{array} \right.
\end{equation}
An example of a PB potential profile under conditions appropriate
for Laponite is shown in Fig. \ref{Potential_2_discs}.

Once $\varphi({\bf r})$ and the resulting concentration profiles
$\rho^\alpha({\bf r})$ are known, the local stress tensor may be
evaluated at each point of the WS cell
\begin{eqnarray}
\label{eq:stress_tensor}
&&\stackrel{\longleftrightarrow}{\bbox{\Pi}}({\bf r}) = 
p({\bf r})  \stackrel{\leftrightarrow}{\bf I}
\,- \, \frac{\varepsilon}{4\pi} \,\bbox{\nabla}\varphi
\otimes
\bbox{\nabla}\varphi,\\
&& p({\bf r}) =  k_{_{B}}T \sum_{\alpha = +,-}\rho^{\alpha}({\bf r}) +
\frac{\varepsilon}{8\pi} | \bbox{\nabla} \varphi({\bf r}) |^2
\end{eqnarray}
where $\stackrel{\leftrightarrow}{\bf I}$ denotes the unit tensor.
The force acting on platelet $i \in (1,2)$ follows by integrating
the stress tensor (\ref{eq:stress_tensor}) over both sides,
$\Sigma_{{\cal P},i}^{+}$ and $\Sigma_{{\cal P},i}^{-}$,
of the platelet
\begin{equation}
{\bf F}_i = -\int_{\Sigma_{{\cal P},i}^+ ;\, \Sigma_{{\cal P},i}^-} 
\stackrel{\longleftrightarrow}{\bbox{\Pi}}\!\cdot\, {\bf dS}_i.
\label{eq:force}
\end{equation}
In the case under consideration, of two identical coaxial discs, the 
force ${\bf F}_1 = - {\bf F}_2$ is along the unit vector $\hat{\bf z}$
of the cylinder axis. The electric field ${\bf E} 
= -\bbox{\nabla}\varphi$ in the vicinity of platelet $i$ may be
decomposed into continuous and discontinuous parts
\begin{equation}
{\bf E} = {\bf E}_i^{(d)}+{\bf E}_i^{(c)} = 
\pm \frac{2\pi}{\varepsilon} \, \sigma \,
{\bf n}_i + {\bf E}_i^{(c)}.
\label{eq:decomp}
\end{equation}
The force ${\bf F}_i$ may then also be expressed as
\begin{equation}
{\bf F}_i = \sigma \int_{\Sigma_{{\cal P},i}} {\bf E}_i^{(c)} \,d^2 S
\qquad (i=1,2),
\label{eq:forcefin}
\end{equation}
where the integration runs over the surface of platelet $i$. 
The numerical evaluation of the force from (\ref{eq:forcefin})
poses technical difficulties associated with the removal
of the discontinuity suffered by the electric field across
the platelet which requires some care (cf. Appendix \ref{app:b}).

Returning to the definition (\ref{eq:force}), the surface integral
may be transformed using the mechanical equilibrium condition 
\begin{equation}
\bbox{\nabla}\cdot \stackrel{\longleftrightarrow}{\bbox{\Pi}}
({\bf r}) = \bbox{0}
\label{eq:equil}
\end{equation}
into the following integral over the surface $S$, enclosing the
upper half of the WS cell (cf. Fig. \ref{fig2plates})
\begin{equation}
{\bf F}_1 = - {\bf F}_2 = -\int_S
\stackrel{\longleftrightarrow}{\bbox{\Pi}}\!\cdot\, {\bf dS}.
\label{eq:forcebis}
\end{equation}
Because ${\bf n}\!\cdot\!\bbox{\nabla}\varphi = 0$ on the outer
surface $\Sigma$ of the WS cell, and $\partial \varphi({\bf r})/
\partial z = 0$ by symmetry on the cross-section of the cylinder
at $z=0$, the term $\bbox{\nabla}\varphi\otimes\bbox{\nabla}\varphi$
gives a vanishing contribution to the force ${\bf F}_1$,
which reduces to
\begin{equation}
{\bf F}_1 = - \int_S \,p({\bf r})\, {\bf d S}.
\end{equation}
This may be further simplified by noting that 
$\hat{{\bf z}}\!\cdot\!{\bf n}=0$ on the lateral surface
of the cylinder (parallel to the axis), so that finally
\begin{equation}
F_1^z = \int_{z=0}\, p({\bf r})\, d S - \int_{z=h}\, p({\bf r})\, d S.
\label{eq:forastuce}
\end{equation}

The numerical consistency between results based on Eq. 
(\ref{eq:forastuce}) and on Eq. (\ref{eq:forcefin})
(cf. Appendix \ref{app:b}) has been carefully checked. 
Explicit calculations were carried out in the limit 
of vanishing clay concentration ($n \to 0$). This was
achieved by choosing a WS cylindrical cell large enough for
the local electric field to vanish before the outer 
surface of the cell is reached. For platelets with 
the physical characteristics of Laponite ($r_0=150$\AA$\,$
and $Z=1000$), and salt concentrations $n_S 
\stackrel{>}{\sim} 10^{-3}$M, this condition
is met with a cell volume $v = 4 \times 10^8$\AA$^3$
(corresponding to a clay concentration 
$n = 8.3 \times 10^{-6}$M) and an aspect ratio
$h/r_0=1.5$. A test that the WS cell was chosen sufficiently
large is provided by checking that the calculated 
concentration $n_S$ of salt in the cell coincides with the
preset reservoir concentration $n_S'$, as expect for vanishing
clay concentration. The forces calculated within PB theory
are plotted versus the distance between platelets in Fig.
\ref{force}, and compared to the analytical prediction
of LPB theory \cite{REF16}
\begin{equation}
\label{Force_LPB_n0}
F^z(d) = (\pi r_0)^2 \frac{4 \pi \sigma^2}{\epsilon}
\int_0^\infty J^2_1(x) \frac{1}{x} 
\exp \left[ -\frac{d}{r_0} \sqrt{x^2 + \kappa_{_{D}}^2 r_0^2} \right].
\end{equation}
For $Z=5$ the PB and LPB results coincide, but in the physically
relevant case $Z=1000$ shown in Fig. \ref{force},
LPB theory overestimates the force by an order
of magnitude. This finding is consistent with
the overestimation of the absolute value of the potential
around a platelet by LPB theory, as illustrated in
Fig. \ref{Potentials}c. Note that unlike its LPB counterpart,
the PB force does not decay exponentially with the distance
between platelets.

\section{conclusion}
Swollen stacks of monodisperse clay platelets are conveniently modelled
within a Wigner-Seitz cell representation. Relevant electrostatic and 
thermodynamic properties are derived from the inhomogeneous counterion 
and coion concentration profiles related to the local electrostatic 
potential by Poisson's equation. Adopting the {\it primitive} model
point of view for the solvent and neglecting spatial correlations between
macroions leads to the closed PB equation for the potential 
$\varphi({\bf r})$ which is solved subject to Neumann boundary conditions on
the confining surface of the WS cell. Whilst the linearised LPB version 
of the theory yields to analytic treatment \cite{REF15,REF16}, the present 
paper concerned with the numerical solution of the full non-linear PB 
problem and of a hybrid PB/DH version of the theory. Rather than solving 
the two-dimensional non-linear partial differential PB equation on a grid,
it was found that a more adequate and stable method is to reduce the 
problem to a non-linear integral equation for $\varphi({\bf r})$, involving
an electrostatic Green's function satisfying appropriate boundary conditions.
Three equivalent routes are proposed, each involving a Green's function
which may be calculated explicitly in terms of a Bessel-Dini series, similar 
to that used for the solution of the LPB problem.

While the present results generally confirm the qualitative trends 
predicted by the LPB analysis \cite{REF15,REF16}, there are considerable
quantitative differences at physically relevant surface charge densities 
of the Laponite platelets. The main findings may be summarized as follows
\begin{itemize}
\item[{\bf a)}] LPB theory overestimates the magnitude of the local potential
by typically a factor of two for a platelet of $10^3e$.
\item[{\bf b)}] Rescaling of the platelet charge to a lower effective value
to force good agreement of the PB and LPB potentials at the centre of the 
platelet fails because the shapes of the potential profiles differ 
significantly, particularly near edges.
\item[{\bf c)}] The scenario of the variation of various properties with
the WS aspect ratio, for a fixed value of the cell volume, derived from
non-linear PB theory confirms the LPB predictions: the grand potential
(or the Helmholtz free energy) goes through a minimum at a well-defined
aspect ratio $h/r_0$; at this same aspect ratio, the osmotic
pressure is also at its minimum, where it coincides with the monotonically
decreasing disjoining pressure, while the quadrupole moment of the charge 
distribution within the WS cell vanishes. Thus the same equilibrium aspect 
ratio is selected by thermodynamic, mechanical, osmotic and electrostatic criteria.
\item[{\bf d)}] The hybrid PB/DH theory, which is linearised with respect to
edge effects yields rather accurate potential profiles, but rather poor
thermodynamic properties; the agreement with full PB theory is improved
upon re-exponentiation of the Bessel-Dini series for the local potential.
\item[{\bf e)}] The same Green's function methodology allows an accurate
calculation of the force between the two coaxial platelets. The force 
calculated within PB theory is an order of magnitude smaller than the LPB
prediction, and decreases with the distance between the platelets
in a non-exponential fashion, up to distances of the order of the radius 
$r_0$ of the platelets.
\end{itemize}

In future work it is planned to include microion correlations and discrete 
solvent (hydration) effects, by generalising the PB functional
(\ref{functional_helmholtz}), along the lines proposed by Biben 
{\it et al.} \cite{REF3}. Generalisation of the present methodology to
flexible charged membranes would also be helpful in describing 
charged soap films or smectite 
clay particles of lateral dimensions larger than those of Laponite.

\vspace{0.5cm}
{\large\bf Acknowledgments} \\
E.T. thanks J.O. Fossum, G. Manificat, T. Nicolai and 
F. van Wijland for interesting discussions. R.J.F.L. de C. has carried 
out work at the ENS de Lyon as part of a project financed by the European 
Commission through the Training and Mobility of Researchers (TMR) programme. 
He is now at UCL, funded by NERC. The present collaboration was facilitated
by a grant from the British-French Alliance Programme.


\begin{table}
\begin{tabular}{|c|c|c|c|}
      &$Z=5$  &$Z=100$&$Z=1000$\\ 
\hline\hline
\multicolumn{4}{c}{ }\\
\multicolumn{4}{c}{$n_S/n_S'$}\\
\hline
LPB           & 0.975 & 0.648 & 0.233  \\
PB/DH         & 0.976 & 0.720 & 0.564  \\
PB            & 0.976 & 0.725 & 0.574  \\
\hline\hline
\multicolumn{4}{c}{ }\\
\multicolumn{4}{c}{$n_S/10^{-3}$M}\\
\hline
LPB and PB    & 4.878 & 3.623 & 2.870 \\
PB/DH         & 4.881 & 3.602 & 2.820 \\
\end{tabular}
\caption{Values of the Donnan ratios $n_s/n_s'$ obtained 
within LPB, PB/DH or PB theory, for an aqueous solution of
clay concentration $n=5\times10^{-5}$M. The temperature 
is 300K and the radius of the discs $r_0 = 150$\AA. The
reservoir salt concentration in PB and PB/DH is
$n_s'=5\times 10^{-3}$M. The
concentrations of added salt in the WS cell are
also indicated.}
\label{table1}
\end{table}

\newpage
{\appendix
\section{}
\label{app:a}
For the three routes outlined in section \ref{sec:green}, $G$ ($G^{\cal B}$ in case 
{\bf b} or $G^\kappa$ in case {\bf c}) is expanded in a Bessel-Dini series
\begin{equation}
G({\bf r},{\bf r}') = \sum_{n \geq 1} G_n(\phi,z, {\bf r}')\, 
J_0\left( y_n \frac{r}{R} \right).
\label{eq:secbd}
\end{equation}
Note that the coefficients of any expansion
\begin{equation}
f(r) = \sum_{n \geq 1} f_n\, J_0\left( y_n \frac{r}{R} \right)
\end{equation}
can be obtained from the inversion relation
\begin{equation}
f_n = \frac{2}{R^2 J_0^2(y_n)} \, \int_0^R \, r \, f(r) \,
J_0\left( y_n \frac{r}{R} \right) \, d r,
\end{equation}
so that the Dirac distribution inside the WS cell can be cast in the form
\begin{eqnarray}
\delta({\bf r}-{\bf r}') &\equiv& \frac{1}{r}\, \delta(r-r')\, \delta(\phi-\phi')\,
\delta(z-z') \nonumber \\
&=& \sum_{n \geq 1} \frac{2}{R^2 J_0^2(y_n)} \, J_0\left( y_n \frac{r}{R} \right) \,
J_0\left( y_n \frac{r'}{R} \right)\, \delta(\phi-\phi')\, \delta(z-z').
\label{eq:secdelta}
\end{eqnarray}
Consider specifically route {\bf c}.
Subsitution of Eqs. (\ref{eq:secdelta}) and (\ref{eq:secbd})
into (\ref{eq:greenkappa}) yields
\begin{equation}
\frac{d^2 G_n}{d z^2} - \frac{\Lambda_n^2}{R^2} \, G_n = 
\frac{2 J_0(y_n r'/R)}{R^2 J_0^2(y_n)}\,
\delta(\phi-\phi')\,\delta(z-z')
\label{eq:appediff}
\end{equation}
where $\Lambda_n^{-2} = y_n^2/R^2 + \kappa^2$. With the boundary condition
\begin{equation}
\frac{d G_n}{dz}\biggl|_{z=\pm h} = 0,
\end{equation}
the solution reads 
\begin{equation}
G_n(\phi,z,{\bf r}') = {\cal C}_n^{\pm}\, \cosh\left( \frac{h\mp z}{\Lambda_n}\right),
\end{equation}
where the superscripts $+$ and $-$ 
refer to the situations $z>z'$ and $z<z'$ respectively.
The coefficients ${\cal C}_n^\pm$ are conveniently
obtained by writing $G_n$ in the form
\begin{equation}
G_n(\phi,z,{\bf r}') = g_n^+(z,\phi) \, \Theta(z-z') \,+ \,
g_n^-(z,\phi)\, \Theta(z'-z).
\end{equation}
Invoking the identity
\begin{equation}
f(z) \, \frac{d}{dz}\delta(z-z')\,=\, -f'(z)\, \delta(z-z') \,+\,
f(z') \, \frac{d}{dz}\delta(z-z')
\end{equation}
the continuity conditions obeyed by $G_n$ at $z=z'$ follow as
\begin{eqnarray}
&&g_n^+(z') =  g_n^-(z') \nonumber \\
&&\frac{d}{dz}g_n^+(z')-\frac{d}{dz}g_n^-(z') = \frac{2}{R^2\,J_0^2(y_n)}\,
J_0\left(y_n \,\frac{r'}{R}\right)\, \delta(\phi-\phi'),
\end{eqnarray}
leading back to Eq. (\ref{eq:cn})

An example of Green's function associated with method {\bf a} can be obtained
by imposing 
\begin{equation}
\frac{d G_n}{dz}\biggl|_{z=\pm h} = 0, \qquad\hbox{for } n \geq 2
\end{equation}
while the non-Neumann character of $G$ reflects itself in the boundary condition
obeyed by $G_1$, that follows from the resolution of (\ref{eq:appediff}) 
with $\kappa=0$. One finds 
\begin{equation}
G_1(z,\phi,{\bf r}') = \frac{1}{R^2}\, \delta(\phi-\phi') \, |z-z'|
\label{eq:appgmeta}
\end{equation}
whereas the remaining terms $G_n, n\geq 2$ are the same as in (\ref{green})
and (\ref{eq:cn}), with $\kappa=0$. 

Finally, method {\bf b} can be illustrated along similar lines
with the following choice of the background
\begin{equation}
{\cal B} \,  = \, -\,\frac{1}{h R^2} \, \delta(\phi-\phi').
\end{equation}
The above density is such that the expansion (\ref{green}), 
with relation (\ref{eq:cn})
and $\kappa=0$, is unchanged for $n\geq 2$ while the first term is 
now slightly modified with respect to Eq. (\ref{eq:appgmeta}), namely
\begin{equation}
G_1(z,\phi,{\bf r}') = \left[ \frac{1}{R^2} \, |z-z'|\, -\,\frac{1}{2 h R^2}\,
\left(z^2+z'^2 \right)\right]\, \delta(\phi-\phi').
\end{equation}


\section{}
\label{app:b}
From Eq. (\ref{eq:forcefin}), the force acting on platelet 1 
sitting at $z=+z_0$ can be written
\begin{equation}
\label{Force}
F_1^z = - 2 \pi \sigma \int_{0}^{r_0} dr \, r \left[
\left(\frac{\partial \varphi(z,r)}{\partial z}\right)_{z=z_0}
\pm \frac{2 \pi}{\epsilon} \sigma \right],
\end{equation}
where $-$ is for $z=z_0^+$ and $+$ is for $z=z_0^-$.
This equation allows the immediate computation of the force 
from the electrostatic potential. However, it should not be used directly 
since the numerical differentiation of the potential for $z=z_0$ and $0<r<r_0$ 
(where $\partial \varphi({\bf r})/\partial z$ 
is discontinuous) is inaccurate. 

Instead, apply Gauss' theorem to (\ref{Force}) to obtain
\begin{eqnarray}
\label{newforce}
F^z  & = & \sigma \frac{Q_1}{\epsilon}
- 2 \pi \sigma \int_{r_0}^R  dr \, r 
\left( \frac{\partial \varphi(z,r)}{\partial z} \right)_{z=z_0}
- \frac{Ze \sigma}{2 \epsilon}
\nonumber \\
& = & - \sigma \frac{Q_2}{\epsilon}
- 2 \pi \sigma \int_{r_0}^R  dr \, r 
\left( \frac{\partial \varphi(z,r)}{\partial z} \right)_{z=z_0}
+ \frac{Ze \sigma}{2 \epsilon},
\end{eqnarray}
where $Q_1$ is the total charge (excluding the platelet) 
in the WS cell volume limited by $z_0<z<h$ and $Q_2$ is the 
total charge in the volume of the WS cell limited by $0<z<z_0$
\begin{eqnarray}
Q_1 & = & 2 \pi \int_{z_0^+}^h dz \int_0^R dr \, r \,\rho_c({\bf r}) \\
Q_2 & = & 2 \pi \int_0^{z_0^-} dz \int_0^R dr \, r \,\rho_c({\bf r}).
\end{eqnarray}
Due to the imposed electroneutrality $Q_1$ and $Q_2$ verify
\begin{equation}
Q_1 + Q_2 = eZ.
\end{equation}
Eq. (\ref{newforce}) requires the computation 
of the derivative of the electrostatic potential at 
$z=z_0$ and $r_0<r<R$, where $\partial \varphi({\bf r}) / \partial z$ 
is continuous. The differentiation can be accurately performed by
fitting $\varphi(z,r)$ to a polynomial in $z$, at fixed $r$.

\newpage


\newpage
\begin{figure}
{\large (a)}
\centerline{\psfig{figure=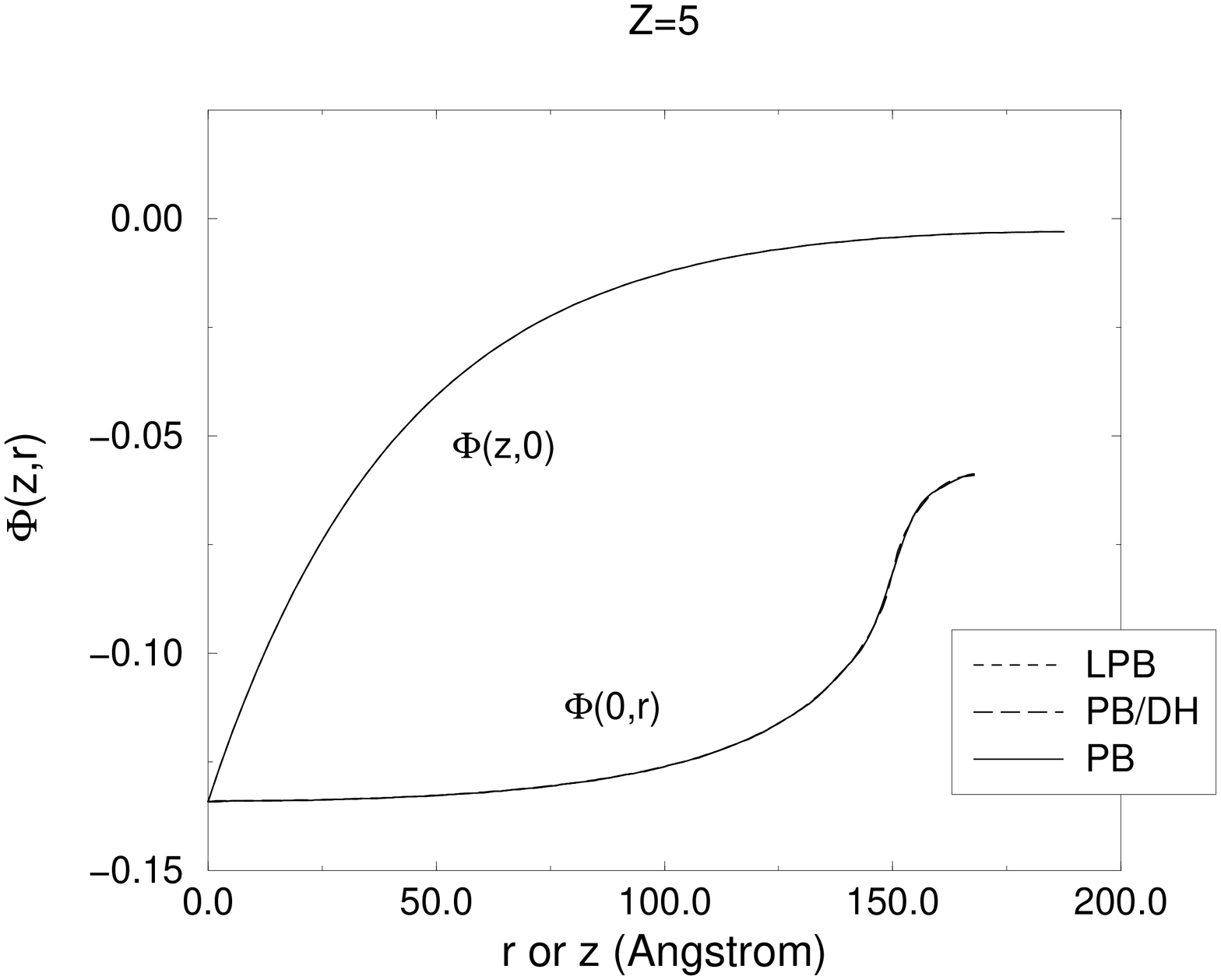,
angle=270,width=7.5cm}}
{\large (b)}
\centerline{\psfig{figure=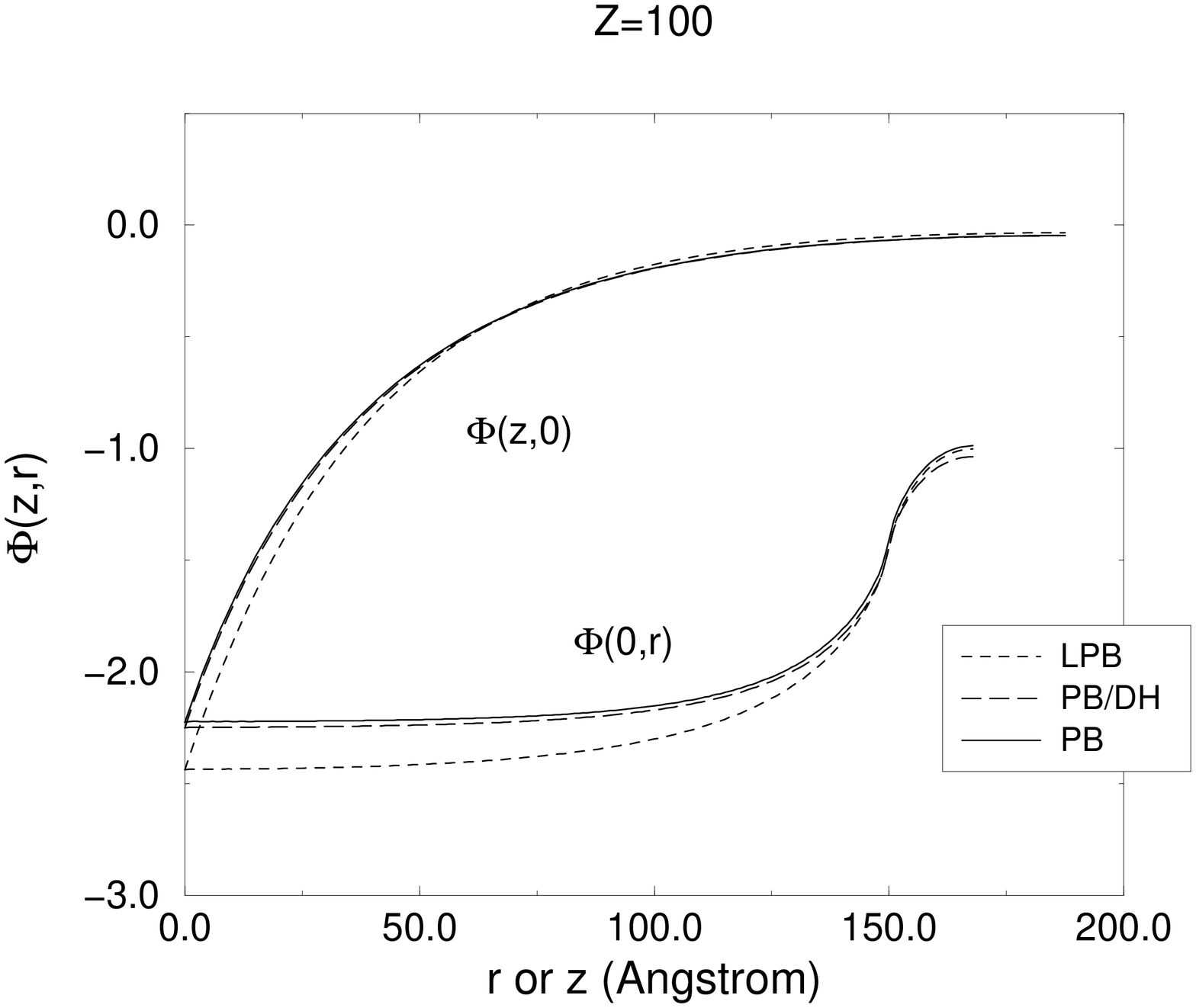,
angle=270,width=7.5cm}}
{\large (c)}
\centerline{\psfig{figure=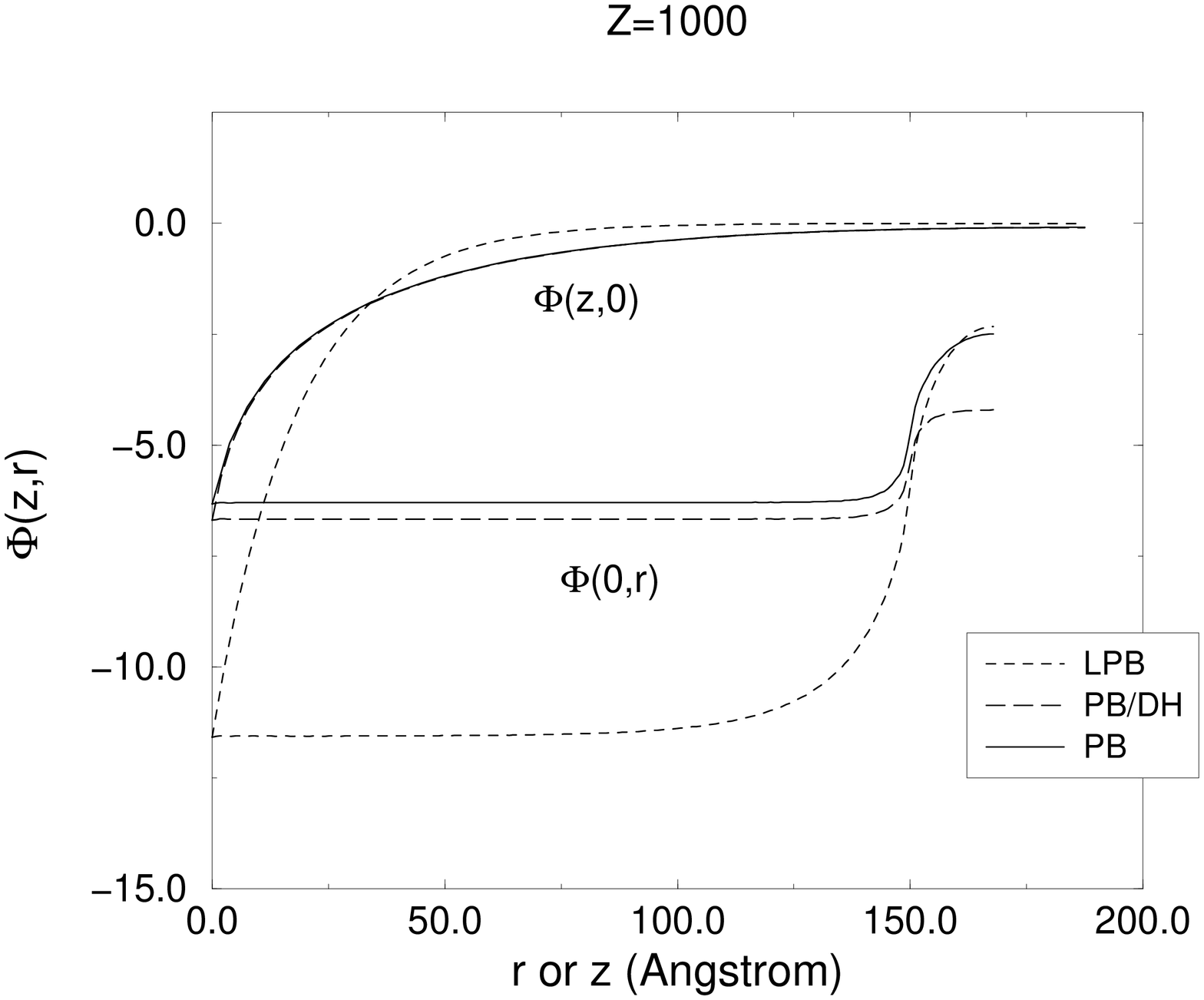,
angle=270,width=7.5cm}}
\caption{\label{Potentials} \protect{\small 
Dimensionless electrostatic potential 
$\Phi(z,r) \equiv \beta e \varphi(z,r)$ 
at $z=0$ versus $r$, and at $r=0$ versus $z$.
The profiles were obtained in LPB (dotted line),
in PB/DH (dashed line) and in the full non-linear 
PB theories (solid line). (See text for details.) The 
surface charge on the clay discs is: 
(a) $Z=5$, (b) $Z=100$ and (c) $Z=1000$.
The concentrations of added salt and the Donnan
ratios $n_S/n_S'$ for these results
are indicated in Table I.}}
\end{figure}

\begin{figure}
\centerline{\psfig{figure=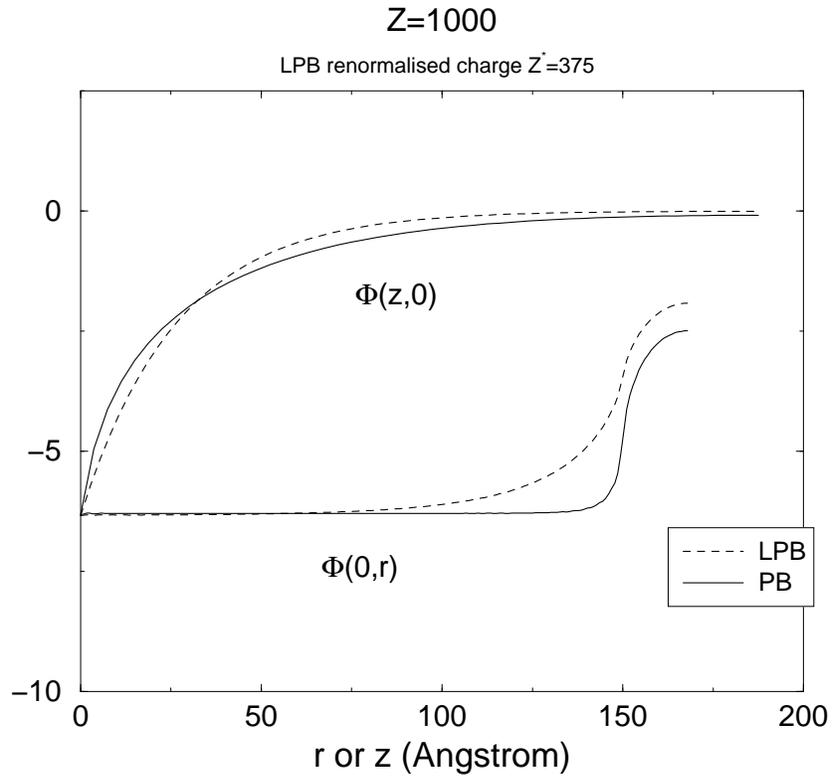,
angle=270,width=11cm}}\vspace{1.5cm}
\caption{\label{fig:renorm} \protect{\small 
Comparison between the non-linear PB potential for $Z=1000$
(as in Figure \protect\ref{Potentials}) and the linearised 
LPB potential, now with an effective renormalised charge 
$Z^*=375$.}}
\end{figure}

\newpage
\begin{figure}
\null\vspace{-1cm}
{\large (a)}
\centerline{\psfig{figure=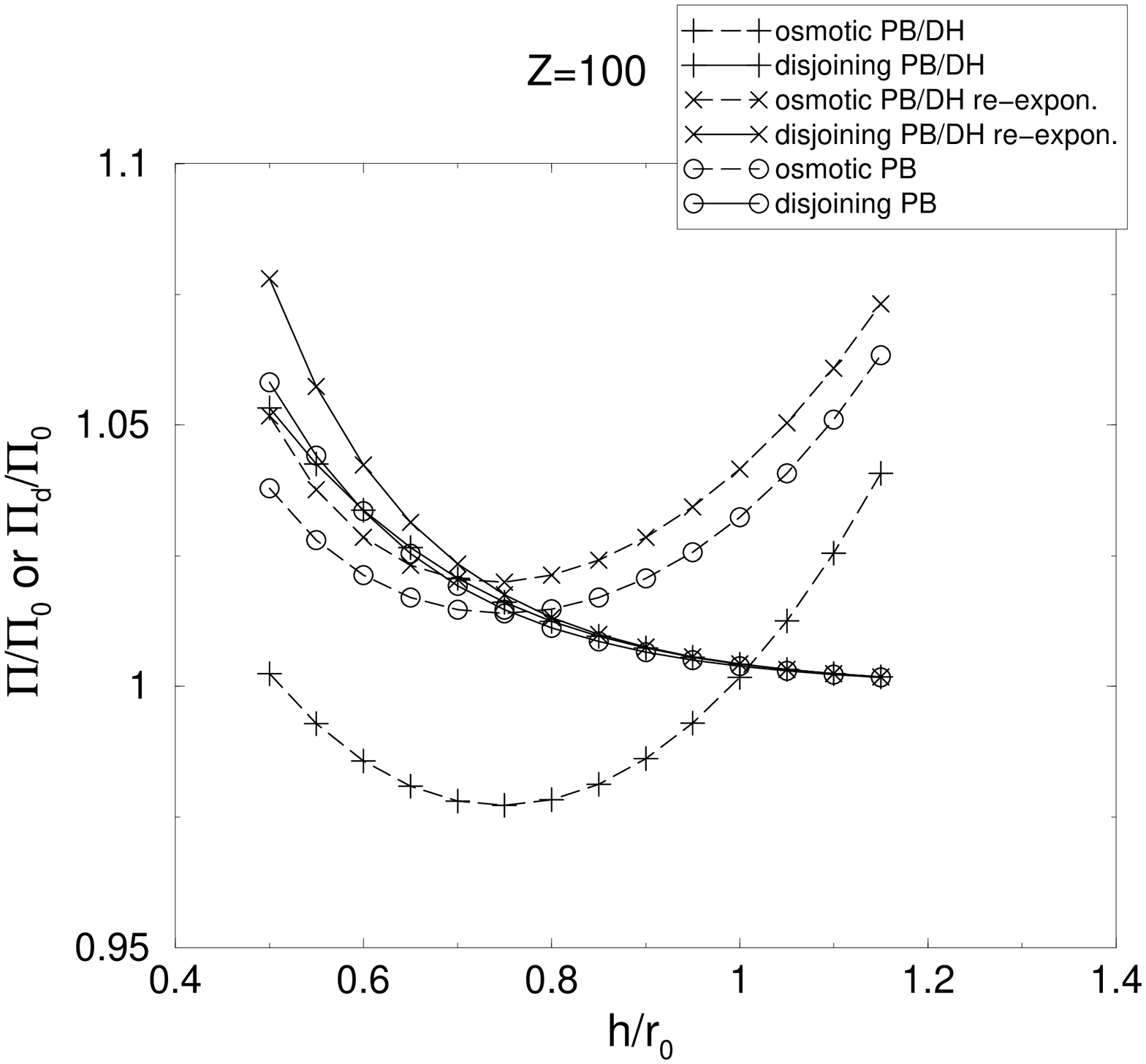,
angle=270,width=7cm}}
{\large (b)}
\centerline{\psfig{figure=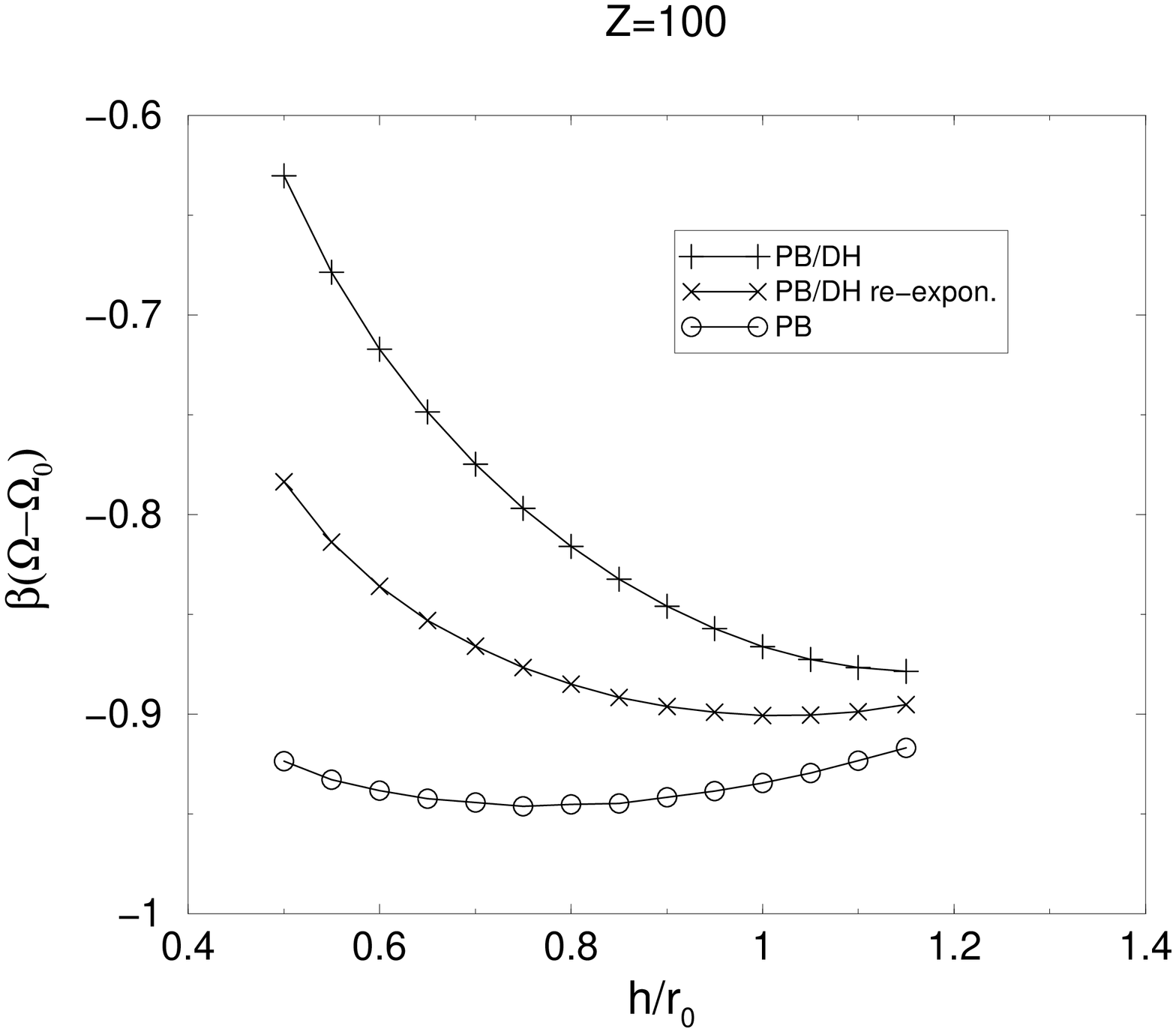,
angle=270,width=7cm}}
{\large (c)}
\centerline{\psfig{figure=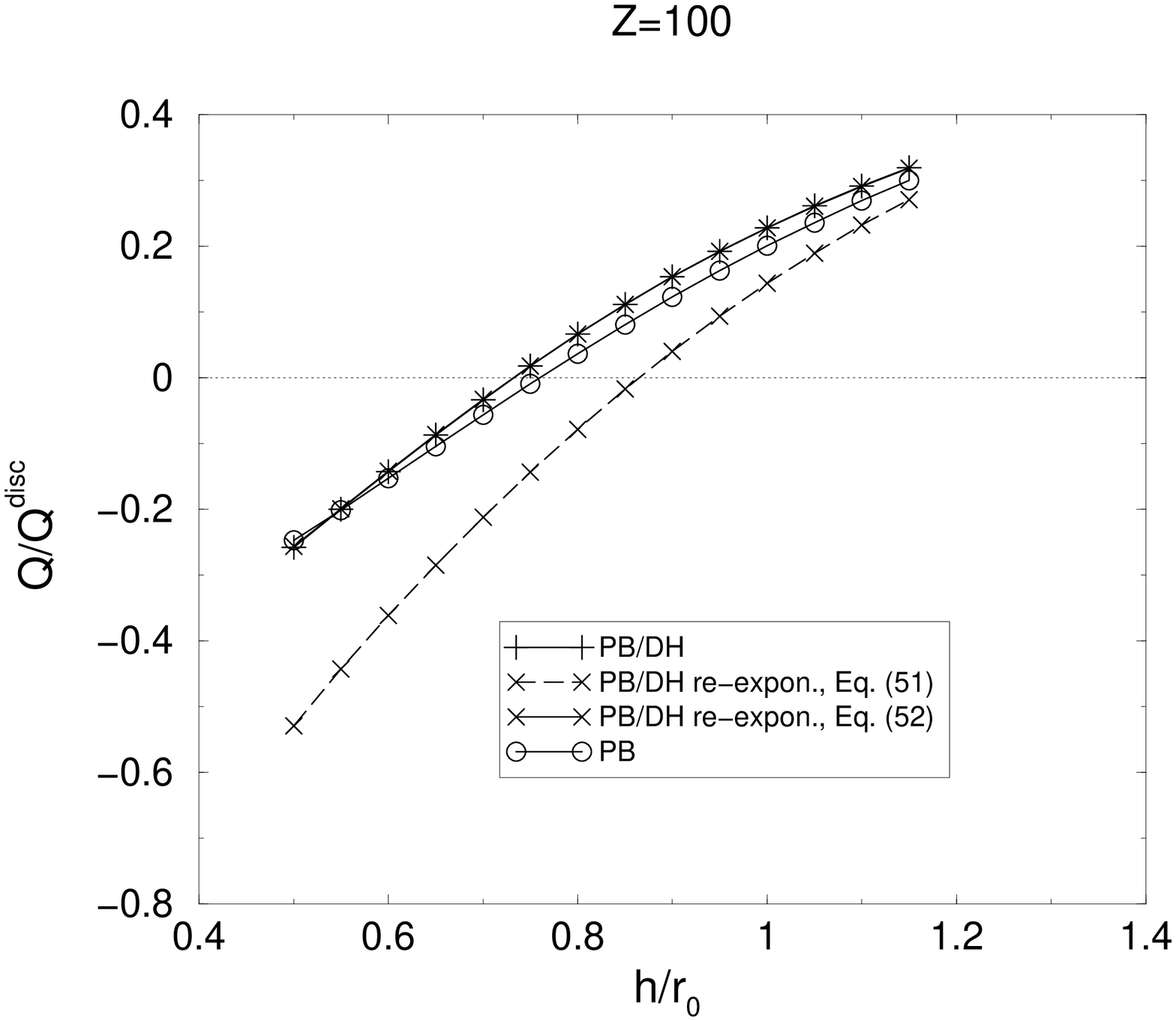,
angle=270,width=7cm}}
\caption{\label{PB.DH_comparisons} \protect{\small 
Results obtained within the PB and PB/DH theories
for an aqueous solution of clay discs
of charge $Z=100$ and radius $r_0 = 150$\AA$\,$.
The clay concentration is
$n = 5 \times 10^{-5}$M and $T=300K$. 
The $\circ$ are PB results, $+$ are
results from PB/DH theory and $\times$ are
from PB/DH with re-exponentiation.
In (a) the osmotic pressure, $\Pi$,
is represented by dashed lines 
and the disjoining pressure, $\Pi_d$, 
by solid lines. (b) shows the grand 
potential while (c) shows the quadrupole moment.
The quadrupole obtained within
PB/DH with re-exponentiation
is not consistent: different results
are obtained from Eqs. (\protect\ref{Q_volume}) 
(solid line) and (\protect\ref{Q_surface}) 
(dashed line).}}
\end{figure}

\begin{figure}
{\large (a)}
\centerline{\psfig{figure=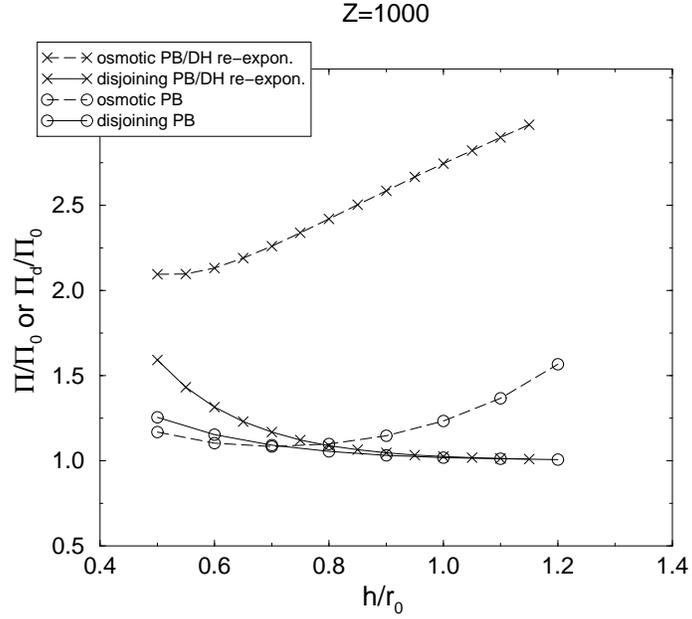,
angle=270,width=9cm}}
{\large (b)}
\centerline{\psfig{figure=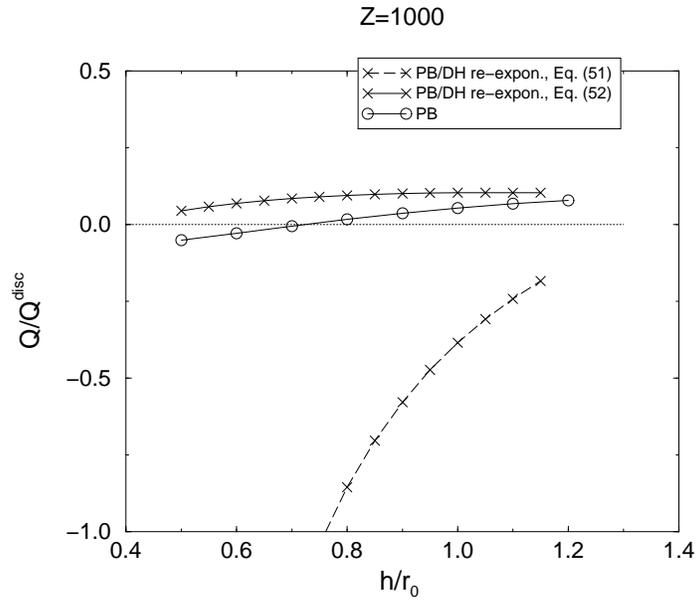,
angle=270,width=9cm}}
\vspace{1cm}
\caption{\label{PB.DH_vs_PB} \protect{\small 
Results obtained within the PB and PB/DH theory 
with re-exponentiation for an aqueous solution of 
clay discs of charge $Z=1000$ and radius 
$r_0 = 150$\AA$\,$. The clay concentration is
$n = 5 \times 10^{-5}$M and $T=300K$. 
The $\circ$ are PB results
and $\times$ are from PB/DH with re-exponentiation.
In (a) the solid lines are disjoining pressures
and the dashed lines are osmotic pressures.
(b) shows the quadrupole moment. Different results
for the quadrupole are obtained within PB/DH with 
re-exponentiation, from Eqs. (\protect\ref{Q_volume}) 
(solid line) and 
(\protect\ref{Q_surface}) (dashed line).}}
\end{figure}

\newpage
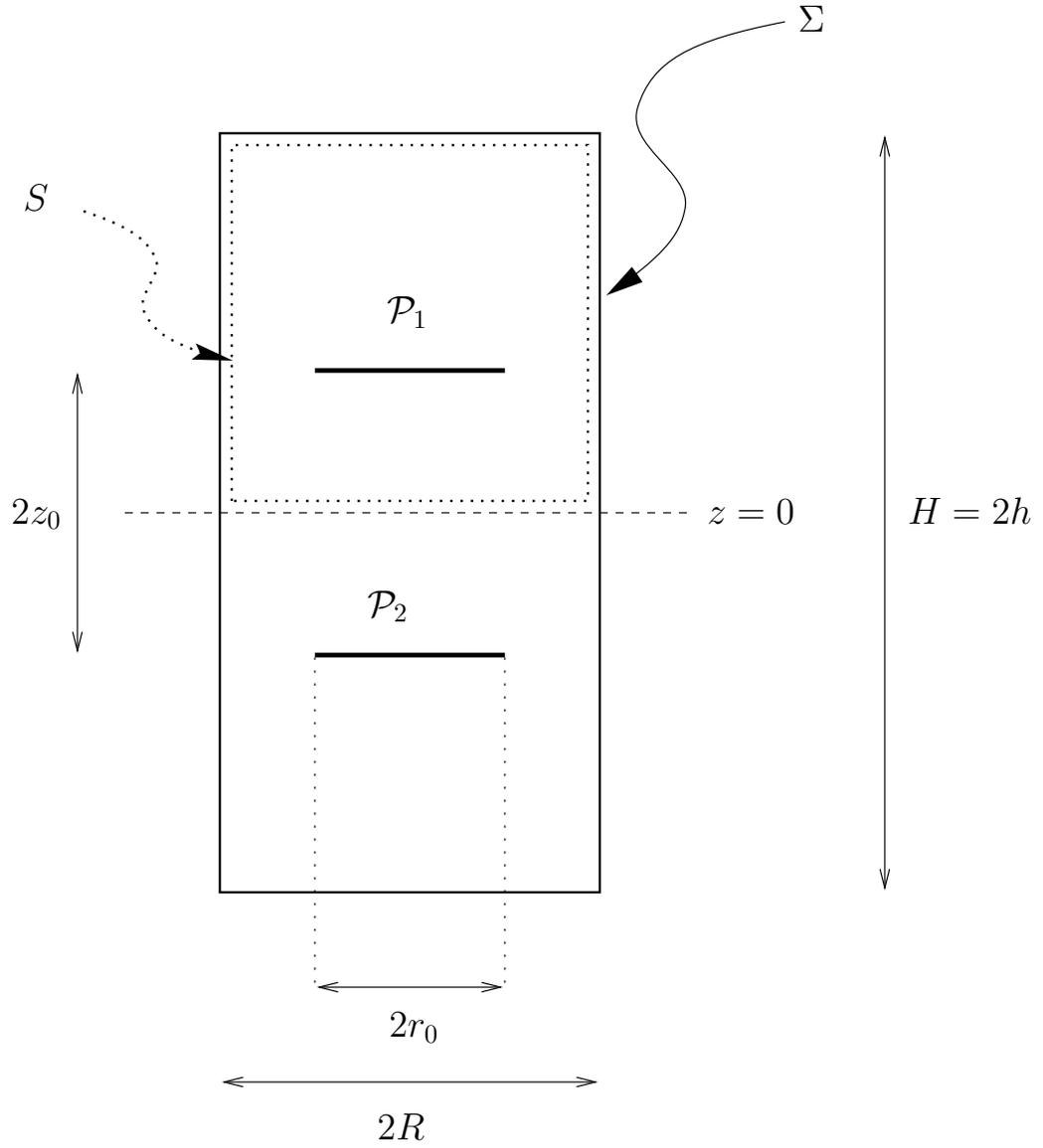
\begin{figure}
$$\input{fig2plates.pstex_t}$$
\caption{\label{fig2plates} \protect{\small Side view of the Wigner-Seitz
cylinder associated with the 2 platelets problem}}
\end{figure}

\newpage
\begin{figure}
\centerline{\psfig{figure=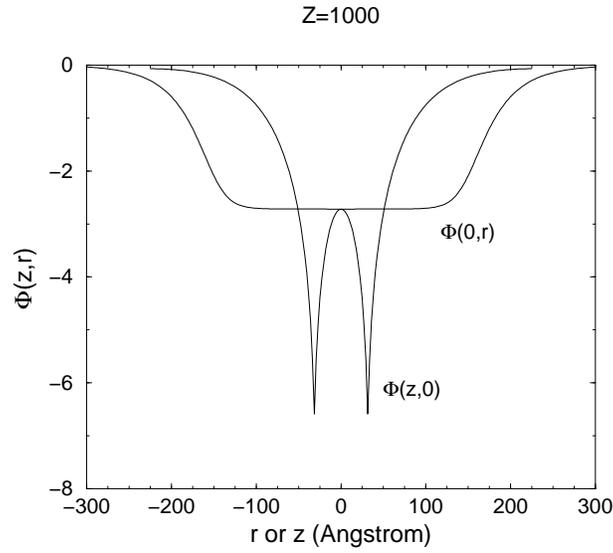,
angle=270,width=8cm}}
\vspace{1.5cm}
\caption{\label{Potential_2_discs} \protect{\small 
Dimensionless electrostatic potential $\Phi(z,r) \equiv \beta 
e \varphi(z,r)$ at $z=0$ versus $r$ and at $r=0$ versus $z$. 
The profiles were obtained by solving the PB equation for an 
aqueous solution of two Laponite discs ($Z=1000$, $r_0=150$\AA) 
at a temperature  $T = 300$K. The distance between the two discs
is $d=0.24r_0$, the height of the cylindrical WS cell is $h=1.5r_0$ 
and the aspect ratio $h/R=0.423$. The concentration of monovalent 
added salt in the WS cell is $n_S = 4.64 \times 10^{-3}$M.}}
\end{figure}

\begin{figure}
\centerline{\psfig{figure=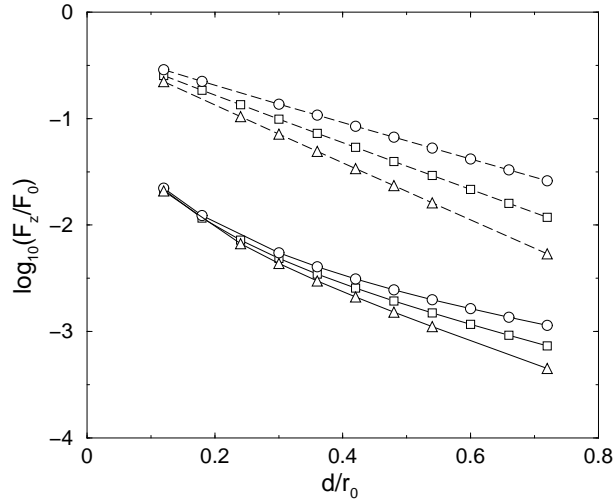,
angle=270,width=8cm}}\vspace{1.5cm}
\caption{\label{force} \protect{\small 
The force between two parallel platelets of Laponite as a function
of the distance separating their coaxial centres ($F_0=4 \pi^2 r_0^2
\sigma^2 / \epsilon$). The solid lines are results obtained 
from non-linear PB and the dashed lines from LPB.
The $\bigcirc$ are for  $n_S = 0.00078$M,
$\Box$ for $n_S = 0.0046$M, and the
$\triangle$ for $n_S = 0.0095$M.}}
\end{figure}

\end{document}

%% file: fig2plates.pstex_t
\begin{picture}(0,0)%
\epsfig{file=fig2plates.pstex}%
\end{picture}%
\setlength{\unitlength}{0.00083300in}%
\begingroup\makeatletter\ifx\SetFigFont\undefined%
\gdef\SetFigFont#1#2#3#4#5{%
  \reset@font\fontsize{#1}{#2pt}%
  \fontfamily{#3}\fontseries{#4}\fontshape{#5}%
  \selectfont}%
\fi\endgroup%
\begin{picture}(5625,7191)(1096,-7345)
\put(1171,-1471){\makebox(0,0)[lb]{\smash{\SetFigFont{14}{16.8}{\rmdefault}{\mddefault}{\updefault}$S$}}}
\put(6031,-346){\makebox(0,0)[lb]{\smash{\SetFigFont{14}{16.8}{\rmdefault}{\mddefault}{\updefault}$\Sigma$}}}
\put(6721,-3436){\makebox(0,0)[lb]{\smash{\SetFigFont{14}{16.8}{\rmdefault}{\mddefault}{\updefault}$H=2h$}}}
\put(5461,-3436){\makebox(0,0)[lb]{\smash{\SetFigFont{14}{16.8}{\rmdefault}{\mddefault}{\updefault}$z=0$}}}
\put(1096,-3436){\makebox(0,0)[lb]{\smash{\SetFigFont{14}{16.8}{\rmdefault}{\mddefault}{\updefault}$2 z_0$}}}
\put(3466,-6661){\makebox(0,0)[lb]{\smash{\SetFigFont{14}{16.8}{\rmdefault}{\mddefault}{\updefault}$2 r_0$}}}
\put(3391,-7306){\makebox(0,0)[lb]{\smash{\SetFigFont{14}{16.8}{\rmdefault}{\mddefault}{\updefault}$2 R$}}}
\put(3451,-2176){\makebox(0,0)[lb]{\smash{\SetFigFont{14}{16.8}{\rmdefault}{\mddefault}{\updefault}${\cal P}_1$}}}
\put(3331,-4021){\makebox(0,0)[lb]{\smash{\SetFigFont{14}{16.8}{\rmdefault}{\mddefault}{\updefault}${\cal P}_2$}}}
\end{picture}